\documentclass[pra,twocolumn,superscriptaddress,  showpacs,footinbib]{revtex4-1}
\usepackage{amsmath}
\usepackage{amssymb}
\usepackage{mathtools}
\usepackage{braket}

\usepackage{txfonts}

\usepackage{graphicx}
\usepackage{xcolor}

\usepackage[colorlinks,urlcolor=blue,linkcolor=blue,citecolor=blue]{hyperref}

\usepackage{footmisc}

\usepackage{dsfont}

\usepackage{siunitx}
\usepackage{calc}
\usepackage{comment}
\usepackage[normalem]{ulem}

\DeclarePairedDelimiter{\abs}{\lvert}{\rvert}
\newcommand{\de}{\mathrm{d}}
\newcommand{\uImm}{\mathrm{i}}
\newcommand{\nepero}{\text{e}}
\newcommand{\Li}[1]{\mathrm{Li}_{#1}\!}
\renewcommand{\Re}{\mathop{\mathrm{Re}}}
\renewcommand{\Im}{\mathop{\text{Im}}\nolimits}%
\newcommand{\beq}{\begin{equation}}
\newcommand{\eeq}{\end{equation}}
\graphicspath{{figures/}}

\begin{document}

\title{Long-range Ising and Kitaev Models: Phases, Correlations and Edge Modes}

\author{Davide Vodola}\thanks{These two authors contributed equally to the present work.}
\affiliation{icFRC, IPCMS (UMR 7504) and ISIS (UMR 7006), Universit\'{e} de Strasbourg and CNRS, 67000 Strasbourg, France.}

\author{Luca Lepori}\thanks{These two authors contributed equally to the present work.}
\affiliation{icFRC, IPCMS (UMR 7504) and ISIS (UMR 7006), Universit\'{e} de Strasbourg and CNRS, 67000 Strasbourg, France.}
\affiliation{Dipartimento di Fisica e Astronomia, Universit\`{a} di Padova, Via Marzolo 8, 35131 Padova, Italy.}

\author{Elisa Ercolessi}
\affiliation{Dipartimento di Fisica e Astronomia, Universit\`{a} di Bologna and INFN, Via Irnerio 46, 40127 Bologna, Italy.}

\author{Guido Pupillo}\email{pupillo@unistra.fr}
\affiliation{icFRC, IPCMS (UMR 7504) and ISIS (UMR 7006), Universit\'{e} de Strasbourg and CNRS, 67000 Strasbourg, France.}

\begin{abstract}
We analyze the quantum phases, correlation functions and  edge modes for a class of spin-1/2 and fermionic models related to the one-dimensional Ising chain in the presence of a transverse field. These models are the Ising chain with anti-ferromagnetic long-range  interactions that decay  with distance~$r$ as $1/r^\alpha$, as well as a related class of fermionic Hamiltonians that generalise the Kitaev chain, where both the hopping and pairing terms are long-range and their relative strength can be varied. For these models, we provide the phase diagram for all exponents $\alpha$, based on an analysis of the entanglement entropy, the decay of correlation functions, and the  edge modes in the case of open chains. We demonstrate that violations of the area law can occur for $\alpha \lesssim1$, while connected correlation functions can decay with a hybrid exponential and power-law behaviour, with a power that is $\alpha$-dependent. Interestingly, for the fermionic models we provide an exact analytical derivation for the decay of the correlation functions at every $\alpha$. Along the critical lines, for all models breaking of conformal symmetry is argued at low enough $\alpha$.
For the fermionic models we show that the edge modes, massless for $\alpha \gtrsim 1$, can acquire  a mass for $\alpha < 1$.  The mass of these  modes  can be tuned by varying the relative strength of the kinetic and pairing terms in the Hamiltonian. Interestingly, for the Ising chain a similar edge localization appears for  the first and second excited states on the paramagnetic side of the phase diagram, where edge modes are not expected. We argue that, at least for the fermionic chains,  these massive states correspond to the appearance of new phases, notably approached via quantum phase transitions without mass gap closure. Finally, we discuss the possibility to detect some of these effects in experiments with cold trapped ions.

\end{abstract}


\maketitle

\section{Introduction}
Topological superconductors and insulators have generated enormous interest in recent years as they correspond to examples of novel quantum phases that are not captured by the familiar Ginzburg-Landau theory of phase transitions. Breakthrough experiments have already led to the observation of symmetry protected topological phases both in condensed-matter systems~\cite{Hsieh2008} and atomic, molecular, and optical physics~\cite{Hafezi2013,Jotzu2014}. While topological phases are finding applications in fields as diverse as photonics and spintronics, the recent probable observation of Majorana modes~\cite{Mourik2012, Franz2013, Nadj-Perge2014, Deng2012, Das2012, Rokhinson2012, Finck2013} in solid-state materials represents the first major step towards the realization of topological quantum computing.

Majorana modes are non dispersive states with zero energy.
In Ref.~\cite{Kitaev2001}, Kitaev has shown that these modes can exist  localized at the edges of a one-dimensional superconductor made of spinless fermions with short-range (SR) $p$-wave pairing. This model is solvable and the underlying lattice Hamiltonian can be mapped exactly onto the well-known Ising chain in a transverse field in one dimension. For SR interactions, the latter is a text-book  example of Hamiltonian displaying a quantum phase transition, here from an ordered (anti-)ferromagnetic phase to a disordered paramagnetic one. Following earlier theoretical works~\cite{Deng2005, Schneider2012}, recent experiments with cold trapped ions have generated enormous interest by demonstrating that long-range (LR) Ising-type Hamiltonians arise as the effective description for the dynamics of the internal states of cold trapped ions, acting as (pseudo-)spins with two or, recently three, internal states. In these experiments, effective spin interactions are generated by a laser-induced manipulation of the vibrational modes of the ion chain~\cite{Friedenauer2008,Britton2012, Jurcevic2014, Schneider2012, Bermudez2013}, which are naturally long ranged. The resulting Ising-type interactions are antiferromagnetic and decay with distance $r$ as a power-law $1/r^\alpha$, with an {\it adjustable} exponent $\alpha$ usually in the range $0\lesssim \alpha \lesssim 3.5$.

In experiments with cold ions, the quantum state of individual particles can be prepared and measured. As a result, both the static and dynamical properties of the many-body system are accessible. Recent experiments have led to the observation of instances of interaction-induced frustration~\cite{Islam2013}, non-local propagation of correlations~\cite{Hauke2013,Richerme2014,Gong2014, Foss-Feig2015, Cevolani2015} and entanglement in a quantum many-body system~\cite{Schachenmayer2013,Jurcevic2014}. Very recently,  spectroscopy experiments have focused on the precise determination of the excited states of LR models~\cite{Jurcevic2015}.

The experimental works described above are based on the understanding of the phase diagram of the Ising-chain in a transverse field, which is known exactly for SR interactions only. In a seminal work~\cite{Koffel2012}, Koffel, Lewenstein and Tagliacozzo have explored the phase diagram of this system with LR interactions in the parameter range $\alpha \gtrsim 0.5$. The results were intriguing: (i) The connected correlation functions decay with a power-law tail even within the gapped paramagnetic phase, at odds with conventional wisdom inherited from SR models and consistent with earlier predictions for other quantum models with LR interactions~\cite{Deng2005,Gong2014, Hazzard2014, Foss-Feig2015,Cevolani2015}. Crucially, (ii) the entanglement entropy, usually a constant within gapped phases, seems  to scale logarithmically with the system size within the paramagnetic phase for sufficiently small $\alpha \lesssim 1$ as well as sub-logarithmically for $\alpha > 1$. This is remarkable as would signal a violation of the so-called ``area law'', dictating the behaviour of the entanglement entropy in SR quantum mechanical systems. These studies also confirm (iii) the persistence of antiferromagnetic and paramagnetic orders with decreasing $\alpha$.

Research in the area of topological phases with LR interactions is very active, and several possible experimental realisations  have been recently proposed. In particular, Kitaev chains with non-local hopping and pairing may be realized in solid state architectures with so-called helical Shiba chains, made of magnetic impurities on an $s$-wave superconductor~\cite{Pientka2013,Pientka2014}. For atomic and molecular systems, key implementations of topological phases have been proposed with polar molecules, dipolar ground state atomic quantum gases and Rydberg excited atoms~\cite{Duan2003, Garcia-Ripoll2004, Baranov2005, Cooper2005, Micheli2006, Brennen2007, Lahaye2009, Cooper2009,  Weimer2010, Levinsen2011,Baranov2012,Yao2012, Peter2013, Yao2013, Manmana2013, Gorshkov2013, Maghrebi2015, Yao2015}. In addition, the famous Haldane phase may be soon realised in cold ion experiments with three internal states per ion~\cite{Cohen2014,Senko2015}, simulating spin-1 particles. For this latter model, very recent theoretical work~\cite{Gong2015} has demonstrated that major features of symmetry-protected topological order  can persist for LR interactions. 
It remains an open question to determine the validity of these results for generic symmetry-protected topological phases with LR interactions.\\

In this work, we analyze the quantum phases, correlation functions and edge-mode localisation of a class of spin-1/2 and fermionic models related to the one-dimensional Ising chain in the presence of a transverse field. These models are the Ising chain with anti-ferromagnetic LR interactions, as discussed in Ref.~\cite{Koffel2012}, as well as a class of Hamiltonians corresponding to a generalization of the Kitaev chain, where both the hopping and pairing terms are LR with an algebraic decay $1/r^\alpha$, and their relative strength can be varied.

 For these models, we provide the phase diagram for all exponents $\alpha$, based on an analysis of: (i) the entanglement entropy; (ii) the decay of correlation functions in all phases;   (iii)  the mass and the localization properties of the edge modes when the chains are open.
 
 In the case of the long-range Ising (LRI) chain we utilize numerical calculations based on the density-matrix-renormalization group (DMRG) method~\cite{White1992, Schollwock2005}, while the long-range Kitaev-type (LRK) models remain exactly solvable for all $\alpha$, allowing for analytical calculation.\\

The following results are obtained for all models: 

(i) A violation of the area law for the entanglement entropy occurs in gapped regions with $\alpha\lesssim 1$. For $\alpha\gtrsim 1$, no violation is found. 

(ii) For {\it any finite $\alpha$}, connected correlation functions within the gapped phases display a hybrid decay that is exponential at short distances and algebraic at long ones. The power of the algebraic decay, however, as well as the extension of the two decay regimes, depends  on $\alpha$. However, when $\alpha\lesssim 1$, the connected correlation functions show a purely algebraic decay.

(iii) For the LRK models, we provide an {\it exact analytical} expression for the decay of correlation functions within the gapped phases that describes the hybrid behaviour with distance mentioned above and explains its origin. 

(iv) Along the critical lines, we demonstrate that conformal symmetry is broken for sufficiently small $\alpha$, by analyzing the finite size scalings of the Von Neumann entropy and of the energy density for the ground states, as well as the behaviour of the dynamical exponent with $\alpha$ around the minima of the spectrum.

(v) We find the existence of two kinds of edge modes: {\it massless} and {\it massive}. For the LRK models, massless (Majorana) modes, as previously found in~\cite{Vodola2014}, appear in the antiferromagnetic region of the phase diagram for large $\alpha$. The antiferromagnetic phase for the LRK models is defined in analogy with that of the short-range Ising chain. The massive modes, instead, are entirely new and are found to appear in a broad area of the antiferromagnetic phase for $\alpha\lesssim 1$ and when we choose the unbalance $\epsilon$ between the strengths of the hopping and pairing terms to be different from~1. This results suggests, for $\alpha\lesssim 1$, a restoration of the $\mathds{Z}_2$ symmetry associated with the (absence of) ground state degeneracy, as well as a possible transition to a novel $\mathds{Z}_2$ symmetric phase and  without mass gap closure. Interestingly, if we choose $\epsilon=1$, the massless modes survive for all $\alpha>0$ and they are  exponentially localized at the edge of the system.


(vi) For the  LRI chain in the antiferromagnetic phase, edge modes  are massless for all $\alpha$, and,  up to numerical precision are exponentially localized at the edge of the chain, in contrast, e.g.\@ to \cite{Gong2015}.  However, surprisingly we find  in the paramagnetic phase a  localization of {\it excited, gapped, energy eigenstates} for $\alpha \lesssim 1$, which for $\alpha \gtrsim 1$ are instead delocalized in the bulk.  

(vii) We finally discuss the persistence of some of the LR effects discussed above (e.g., hybrid decay of correlation functions and edge mode localization) in small chains of up to 30 sites, as relevant to current experiments.\\

The paper is organized as follows. In Sec.~\ref{sec:Hamiltonians} we introduce the model Hamiltonians that we consider in this work [Sec.~\ref{sec:HamA}], the observables that are used to characterize the various phases [Sec.~\ref{sec:HamB}], and present the corresponding phase diagrams [Sec.~\ref{sec:HamC}]. In particular, in Sec.~\ref{sec:HamD} we discuss the critical lines of the Ising and Kitaev models, and argue that conformal symmetry is broken for sufficiently small $\alpha$. In Sec.~\ref{sec:Correlations} we provide an analytic calculation of the  correlation functions for the fermionic Hamiltonians that explains the hybrid exponential and algebraic decay observed in these LR models [Secs.~\ref{sec:CorrA} and~\ref{sec:CorrB}]. In Sec.~\ref{sec:CorrC} we provide a numerical comparison with results for the LRI chain, displaying similar behaviour. In Sec.~\ref{sec:Edges} we analyze the edge modes in the LRI and LRK chains. In particular, in Sec.~\ref{sec:EdgesA} we analyze the  properties of gapless Majorana modes that are found in the anti-ferromagnetic phases of the LRK models for $\alpha\gtrsim 1$. In Sec.~\ref{sec:EdgesB}, instead, we demonstrate that the edge modes can become massive for $\alpha\lesssim 1$, signalling a transition to a new phase. We discuss similar results obtained for some   excited  states that get localized on the edges in the paramagnetic phase of the LRI model for $\alpha \lesssim 1$. In Sec.~\ref{sec:CorrD} we discuss the observability of some of the results above in small chains of a few tens of particles, as relevant for cold ions experiments. Finally, Section~\ref{Conclusions} discusses conclusions and outlook.

\section{Model Hamiltonians and quantum phases}\label{sec:Hamiltonians}
In this section we introduce the model Hamiltonians that we consider in this work and present the corresponding phase diagrams that we compute based on results from the entanglement entropy, decay of correlation functions, spectrum of excitations, and edge mode localization, as discussed in detail in the following sections.

\subsection{Model Hamiltonians}\label{sec:HamA}

\subsubsection{LR Ising model.}
In this work, we are interested in Ising-type Hamiltonians with LR interacting terms. The long-range Ising Hamiltonian~\cite{Koffel2012} reads
\begin{equation}
H_\text{LRI} = \sin \theta \sum_{i=1 ; j>i}^{L} \frac{\sigma^x_i\sigma^x_j}{\abs{i-j}^\alpha} + \cos \theta \sum_{i=1}^L \sigma^z_i \, ,
\label{LRI}
\end{equation}
where $\sigma_j^\nu$ ($\nu=x,y,z$) are Pauli matrices for a spin-1/2 at site $j$ on a chain of length $L$. The first term on the right hand side of Eq.~\eqref{LRI} describes spin-spin interactions that we choose antiferromagnetic (AM) with $\sin \theta>0$ (or equivalently $0<\theta<\pi$). The second term describes the coupling of individual spins to an external field pointing in the $z$-direction. Thus, while the first term favors an antiferromagnetic phase with spins pointing along the $x$ direction, the second terms favors a paramagnetic (PM) phase where all spins align along $z$. In the case of SR interactions (i.e., for $\alpha \rightarrow \infty$) the Hamiltonian  Eq.~\eqref{LRI} is exactly solvable and a quantum phase transition between these two phases is known to occur at  $\theta_c=\pi/4$. Reference~\cite{Koffel2012} has shown numerically that a quantum phase transition separating the AM and the PM phases survives for all finite $\alpha \gtrsim 0.5$. Below, we are interested in exploring the phase diagram of Eq.~\eqref{LRI} for all $\alpha$ and $\theta$.

\subsubsection{Long-range Kitaev chains.}
Related to the LRI chain, in the following we introduce and analyze a class of fermionic Hamiltonians of the form
\begin{equation}
H_{\mathrm{LRK}} = \sin \theta \sum_{i,j} \frac{a_i^{\dagger}  a_j + (1+\epsilon) \, a_i  a_j + \mathrm{h. c.}}{\abs{i-j}^\alpha} + 2 \cos \theta \sum_{i} n_i \, ,
\label{double}
\end{equation}
where $a^\dagger_j$ describes the creation operator for a fermionic particle at site $j$, and $n_j=a^\dagger_j a_j$. The Hamiltonians Eq.~\eqref{double} represent generalizations of the Kitaev chain for spinless fermions with superconducting $p$-wave pairing, where both the hopping and pairing terms decay with distance algebraically with exponent~$\alpha$. 
Here, all energies are expressed in dimensionless units and the parameter $\epsilon$ governs the unbalance between the hopping and pairing terms.

In the SR limit $\alpha \to \infty$, Eq.~\eqref{double} maps into the Ising chain Eq.~\eqref{double} via the Jordan-Wigner transformation  \cite{Lieb1961} 
\begin{eqnarray}
\sigma^+_j= a^\dag_j \, \exp\left({\uImm \, \pi \sum_{\ell < j} n_\ell}\right); \\
\sigma_j^z = 2 n_j - 1,
\label{eqn:JordanWigner}
\end{eqnarray}
with $\sigma^+_j = (\sigma_j^x+\uImm \sigma_j^y)/2$. However, at finite $\alpha$ this identification does not hold anymore due to the contributions of the  string operators $\exp\left({\uImm \pi \sum_{\ell < j} n_\ell}\right)$ in Eq.~(\ref{eqn:JordanWigner}). In particular, unlike the LRI, Eq.~\eqref{double} remain exactly solvable, allowing for analytic solutions at any finite $\alpha$.\\


\subsection{Observables}\label{sec:HamB}

\subsubsection{von Neumann Entropy.}\label{sec:entropy} 
Entanglement measures are routinely used to characterize the critical properties of strongly correlated quantum many-body systems~\cite{Amico2008}. A key example is the von Neumann entropy $\mathcal{S}_{\ell}$ that we employ in this work. For a system of $L$ sites that is partitioned into two disjoint subsystems $A$ and $B$ containing $\ell$ and $L-\ell$ sites, respectively, $\mathcal{S}_{\ell}$ is defined as
\begin{equation}
\mathcal{S}_{\ell}=-\mathrm{Tr\,} \rho_{\ell} \log_2\rho_{\ell},
\end{equation} 
where $\rho_{\ell}$ is the reduced density matrix of the subsystem $A$.

Two general behaviors of $\mathcal{S}_{\ell}$ are known for the ground states of one-dimensional SR interacting systems. Within gapped phases, $\mathcal{S}_{\ell}$ saturates to a constant value independently of $\ell$ and thus obeys to the so-called area law~\cite{Eisert2010}. 
On the contrary, $\mathcal{S}_{\ell}$ diverges with $\ell$ for critical gapless phases and, for conformally invariant systems, satisfies the universal law \cite{Calabrese2004}:
\begin{equation}\label{vN}
\mathcal{S}_{\ell}=a+\frac{c}{6}\log_2 d(\ell,L),
\end{equation}
with $d(\ell,L)=(2L/\pi) \sin(\pi \ell/L)$ for the case of open boundaries. Here, $a$ is a non-universal constant and $c$ is the central charge of the theory. The latter characterizes the universality class of the gapless phase~\cite{diFrancesco1997,Henkel1999}. $\mathcal{S}_{\ell}$, and thus the central charge, can in principle be directly computed by numerical techniques, such as Density Matrix Renormalization Group~\cite{White1992, Schollwock2005} as well as by means of analytical methods for quadratic Hamiltonians \cite{Peschel1989,Peschel1999,Peschel2002,Peschel2012}.

In the case of LR models it has been shown~\cite{Eisert2010,Koffel2012,Vodola2014} that the divergence of $\mathcal{S}_{\ell}$ in Eq.~\eqref{vN} can also occur for gapped phases, corresponding to a so-called violation of the area law. Since this violation is found to be logarithmic, an {\it effective} central charge $c_{\rm eff}$ may be defined also within the gapped phases and used to characterize the main features of the phase diagram for LR interactions~\cite{Koffel2012}.

\subsubsection{Correlation functions.}\label{sec:Correlation}
The various quantum phases can be characterized by the decay of two-point correlation functions with distance. 
For the LRI model, we are interested in the connected correlations 
\begin{equation}\label{corr:Ising}
C^{\nu\nu}_{i,j} = \braket{ \sigma^\nu_i \sigma^\nu_j}-\braket{ \sigma^\nu_i }\braket{\sigma^\nu_j},
\end{equation} 
with $\nu=x,z$. 

For the LRK models we are interested both in the density-density correlation function
\begin{equation}\label{eqn:densdens}
g_2(i,j) = \braket{n_i n_j} -\braket{n_i}\braket{n_j}  \,
\end{equation}
and the function
\begin{equation}\label{eqn:string}
\Sigma(i,j)=\braket{(a^\dag_i+a_i)\exp\left(\pi\uImm \sum_{\ell=i+1}^{j-1}a^\dag_\ell a_\ell\right)(a^\dag_j+a_j)}
\end{equation}
that correspond to the functions $C^{zz}_{i,j}$ and $\braket{ \sigma^x_i \sigma^x_j}$ in the LRI model, respectively, via the Jordan-Wigner transformation given above~\cite{Lieb1961}. Since the LRK models are quadratic, the functions Eq.~\eqref{eqn:densdens} and~\eqref{eqn:string} can be directly obtained from the one-point correlations $\braket{a^\dag_i a_j}$ and $\braket{a^\dag_i a^\dag_j}$ via Wick's theorem. In particular, one finds 
\begin{equation}
g_2(i,j)=\abs{\braket{a^\dag_i a^\dag_j}}^2 - \abs{\braket{a^\dag_i a_j}}^2
\end{equation} 
and 
\begin{equation}
\Sigma(i,j) = \det \left(\begin{array}{ccc}
G_{i,i+1} & \dots & G_{i,j} \\
\vdots   & \ddots \\
G_{j-1,i+1} & \dots & G_{j-1,j}
\end{array}\right)
\end{equation}
with $G_{m,n} = \delta_{m n} + 2 \braket{a^\dag_m a^\dag_n}+2 \braket{a^\dag_m a_n}$.\\

For SR interactions, the connected correlations above are known to decay exponentially (algebraically) with distance within the gapped (gapless) phases. Surprisingly, for several models with LR interactions it was reported that algebraic decay of correlations can coexist with an initial exponential decay within gapped phases~\cite{Deng2005,Vodola2014,Foss-Feig2015}. An analytic understanding of this effect has so far proven elusive.\\

In the following we use the decay of correlation functions to characterize the various phases. In particular, for the LRI chain we provide extensive numerical results using DMRG techniques, while for the LRK chains we exploit the integrability of the models to derive an analytic expression of the behaviour of correlation functions in all parameter regimes.
 
\subsubsection{Edge states and edge gaps.}\label{sec:edgeStates}
Localized edge states within topological phases have attracted much interest over the last decade, largely because of possible applications in schemes for topological quantum computing \cite{Nayak2008}. In particular, Ref.~\cite{Kitaev2001} has shown that localized states arise at the edge of a 1D superconductor with $p$-wave SR pairing interactions [described e.g. by the limit $\alpha\to\infty$ of Eq.~\eqref{double}]. The existence of these localized states is related with the spontaneous breaking of the discrete $\mathds{Z}_2$ symmetry associated with the parity of the fermion number: when this $\mathds{Z}_2$ symmetry is broken, two degenerate ground states with different parity appear. Here, they will be labeled by $\ket{0^+}$ and $\ket{0^-}$ in the even and odd parity sectors, respectively.  

As Hamiltonians Eq.~\eqref{LRI} and \eqref{double} are equivalent in the limit $\alpha\to\infty$, the same breaking of $\mathds{Z}_2$ symmetry (now related to spin flips along $x$ direction) described above occurs also for the LRI chain, resulting in the presence of two degenerate edge states in this model. From the discussion above it turns our that the analysis of the edge modes can be utilized to characterize the quantum phases of the system.\\

In this work, we identify the localized states by directly computing their wavefunctions and masses, which can be achieved either numerically for Eq.~\eqref{LRI} using DMRG simulations or exactly for Eq.~\eqref{double}. 
In order to accomplish this task for the LRI model,  it is useful to exploit the Jordan-Wigner transformation Eq.~\eqref{eqn:JordanWigner} to define new fermionic operators $c^\dag_j=\sigma^+_j\exp \left(\uImm \pi \sum_{\ell < j} \sigma^{+}_{\ell} \sigma^{-}_{\ell}\right)$  [similar to Eq.~\eqref{eqn:JordanWigner}] from spin operators $\sigma^{\pm}_j$. We then compute the wave-functions $\varphi^{(1,2)}_j$ of the massless edge modes as~\cite{Stoudenmire2011}
\begin{eqnarray}
\varphi^{(1)}_j = \braket{ 0^{-}| c^{\dagger}_j |0^{+}} +\braket{ 0^{+}| c^{\dagger}_j |0^{-}} \label{eqn:MajoranaLeft} \\
\varphi^{(2)}_j = \uImm \left(\braket{ 0^{-}| c^{\dagger}_j |0^{+}} -\braket{ 0^{+}| c^{\dagger}_j |0^{-}} \right).\label{eqn:MajoranaRight}
\end{eqnarray}
Here, the states $\ket{0^{+}}$ and $\ket{0^{-}}$ are the ground states in the even and odd parity sector also for the LRI model, respectively. The mass of the two modes $\varphi^{(1,2)}_j$, also known as \emph{edge gap} (in order to distinguish it from the usual \emph{mass gap}, i.e.\ the energy difference between the ground state and the first excited bulk state, see Refs.~\cite{Gong2015, Chan2015}), is defined as the difference $E_{\ket{0^{-}}} - E_{\ket{0^{+}}}$, with $E_{\ket{0^{\pm}}}$ being the energy of the state $\ket{0^{\pm}}$.\\

In the following we will also be interested in characterizing the localisation of {\it massive} edge states that are found  in the phase of LRI where the $\mathds{Z}_2$ symmetry is preserved, and thus where a unique ground state $\ket{ 0^{+}}$ occurs in the even-parity sector.
This localization arises for the first two excited states $\ket{1^-}$ and $\ket{2^-}$ for $\alpha\lesssim 1$,  which are instead delocalised in the bulk for $\alpha \gg 1$ [see  Sec.~\ref{sec:EdgesB}]. Their wavefunctions read
\begin{equation}
w^{(1)}_j = \braket{ 1^-| c^{\dagger}_j | 0^+} \qquad w^{(2)}_j = \braket{ 2^-| c^{\dagger}_j | 0^+}.
\label{wavemass}
\end{equation}
Here the mass  of the mode $w^{(s)}_j$ is defined as the difference $\Delta E(s)=E_{\ket{s^{-}}}- E_{\ket{0^{+}}}$ with $s=1,2$.
The difference between Eqs.~\eqref{eqn:MajoranaLeft},~\eqref{eqn:MajoranaRight} and~\eqref{wavemass} is that in the first two expressions the second term on the right hand side of the equations is  nonzero because of the zero-energy condition \cite{Stoudenmire2011}.\\

For the LRK models, the wavefunctions for both the massless and massive modes can be extracted following Ref.~\cite{Lieb1961}. The latter describes a technique for the exact diagonalization of a generic fermionic quadratic Hamiltonian of the form
\begin{equation}\label{eqn:quadraticH}
H= \sum_{i,j=1}^{L} \left[A_{i j} a^\dag_i a_j + \frac{1}{2} \left(a^\dag_i B_{i j} a^\dag_j + \text{H.c.}\right)\right],
\end{equation}
with $A=[A_{i j}]$ and $B=[B_{i j}]$ real matrices. Equation~\eqref{eqn:quadraticH} can be cast in diagonal form as
\begin{equation}
H=\sum_{n=0}^{L-1} \Lambda_n \eta^\dag_n \eta_n. 
\end{equation}
by a singular value decomposition of the matrix $A+B$:
\begin{equation}\label{eqn:SVD}
\Lambda_n = \sum_{ij}\psi_{n j}(A_{j i}+B_{j i}) \phi_{i n}.
\end{equation}
Here, $\Lambda_n$ are single-particle energies (ordered as $\Lambda_0 \leq \Lambda_1 \leq \dots \leq \Lambda_{L-1}$) and $\eta_n$ are fermionic operators defined by the following Bogoliubov transformation
\begin{equation}\label{eqn:Bogoliubov0}
\eta_n = \sum_{j} g_{nj} a_j + h_{nj} a^\dag_j.
\end{equation}
The matrix elements $\phi_{nj}=g_{nj}+h_{nj}$ and $\psi_{nj}=g_{nj}-h_{nj}$ can be directly identified as the wavefunctions of the two Majorana modes $(\eta_{n}+\eta_{n}^\dag)/2$ and $\uImm (\eta_{n}-\eta_{n}^\dag)/2$ with energy $\Lambda_n$, while $g_{nj}$ and $h_{nj}$ are the wavefunctions of $\eta_{n}$ and $\eta_{n}^\dag$.

\begin{figure*}\centering
\includegraphics[width=0.9\textwidth-10pt]{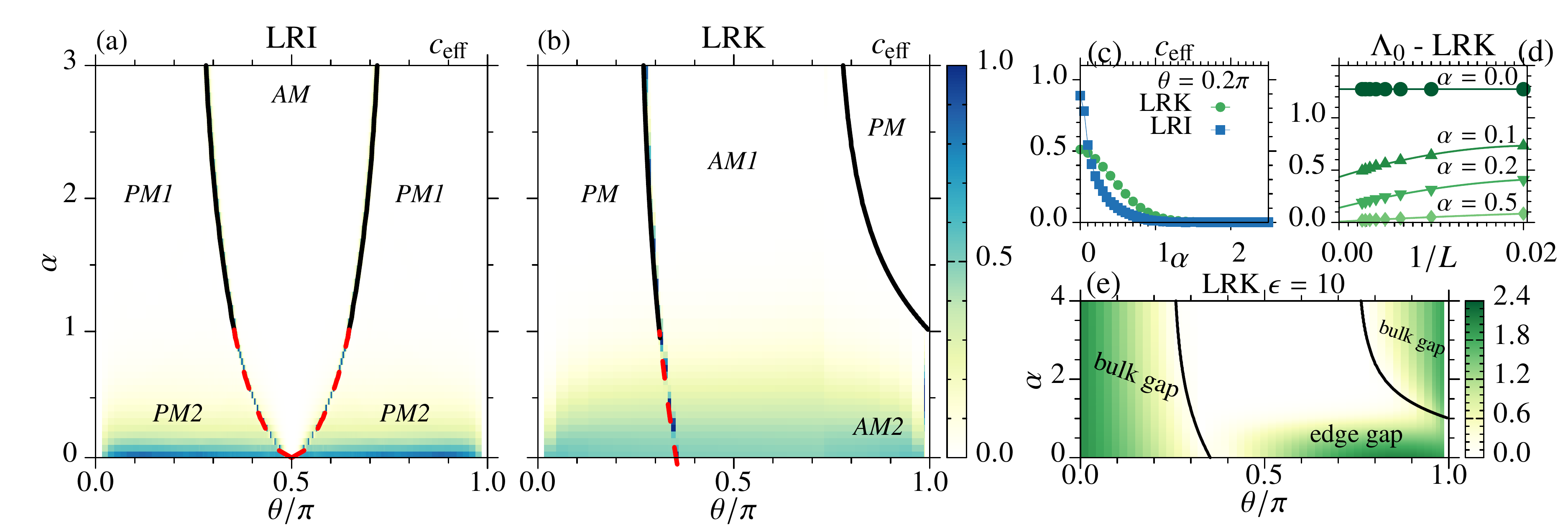}
\caption{(a) Phase diagram of the LRI model Eq.~\eqref{LRI} from the effective central charge $c_\text{eff}$ extracted from the entanglement entropy $\mathcal{S}_{\ell}$ [see Eq.~\eqref{vN}] for a system of $L=100$ sites, as a function of $\alpha$ and~$\theta$. The phase diagram is symmetric with respect to $\theta=\pi/2$.  PM1 and AM denote gapped paramagnetic and antiferromagnetic phases, respectively, with $c_\text{eff}=0$. In the paramagnetic gapped region PM2, $c_\text{eff}>0$ [see panel (c) blue squares] and massive edge modes [see Sec.~\ref{sec:EdgesB} and Fig.~\ref{edgeDLRK}(d)] are found. The continuous black lines correspond to  critical transition lines with central charge $c=1/2$. The dashed red  lines signal an increase of the central charge from 1/2 up to $c \simeq 1$ with decreasing $\alpha$ [see Fig.~\ref{fig:crossingMethod}(b) blue triangles]. (b) Phase diagram of the LRK model Eq.~\eqref{double} from $c_\text{eff}$ extracted from $\mathcal{S}_{\ell}$ in the thermodynamic limit as a function of $\alpha$ and $\theta$, and for $\epsilon=0$. AM1 and PM are gapped phases with $c_\text{eff}=0$ for $\alpha\gtrsim 1$ and $c_\text{eff}\neq 0$ for $\alpha\lesssim 1$ [see panel (c) green circles]. AM1 shows massless edge modes with a hybrid exponential and power-law decay with distance. AM2 shows  edge states with a finite mass $\Lambda_0$,  which increases with decreasing $\alpha$ [see panel (c) and Fig.~\ref{DKLRmassive}(a)]. The continuous black lines correspond to critical transition lines with central charge $c=1/2$. The dashed red  lines signal an increase of the central charge from 1/2 up to $c \simeq 1$ with decreasing $\alpha$.  (c) Effective central charge $c_\text{eff}$ for the LR Ising (blue squares) and the LRK models (green circles) for $\theta=0.2\pi$ as function of $\alpha$.  (d) Scaling with the system size $L$ of the mass $\Lambda_0$ of the edge modes within the AM2 phase, for $\theta=0.75\pi$ and for several $\alpha<1$. (e) Edge gap $\Lambda_0$ of the LRK model for a system of $L=400$ sites and $\epsilon=10$. In the region denoted by ``edge gap'' the $\mathds{Z}_2$ symmetry of the model is restored  and the AM2 phase appears.}
\label{fig1}
\end{figure*}

\subsection{Phase diagrams of LRI and LRK models}\label{sec:HamC}
In this section, we present the phase diagrams for the LR models Eqs.~\eqref{LRI} and~\eqref{double}, obtained from an analysis of the observables described above. The results for the LRI model were obtained numerically via DMRG techniques for chains of a length $L$ up to $L=200$. For all calculations we utilized up to 128 local basis states and 10 finite-size sweeps~\cite{White1992, Schollwock2005}. The discarded error on the sum of the eigenvalues of the reduced density matrix was always less than $10^{-8}$. For the LRK models all results were obtained (semi-)analytically.

\subsubsection{LR Ising model.}\label{sec:phaseIsing}
Our results for the phase diagram of the LRI model are summarized in Fig.~\ref{fig1}(a), where we plot the effective central charge $c_{\rm eff}$ defined in Sec.~\ref{sec:entropy} as a function of the angle $\theta$ and the power $\alpha$ of the antiferromagnetic term in Eq.~\eqref{LRI}. 
Hamiltonian Eq.~\eqref{LRI} is invariant under the transformation $\cos{\theta}\to-\cos{\theta}$ and thus the phase diagram is symmetric around $\theta=\pi/2$.\\

\begin{figure}[t]
\centering
\includegraphics[width=0.5\textwidth-4pt]{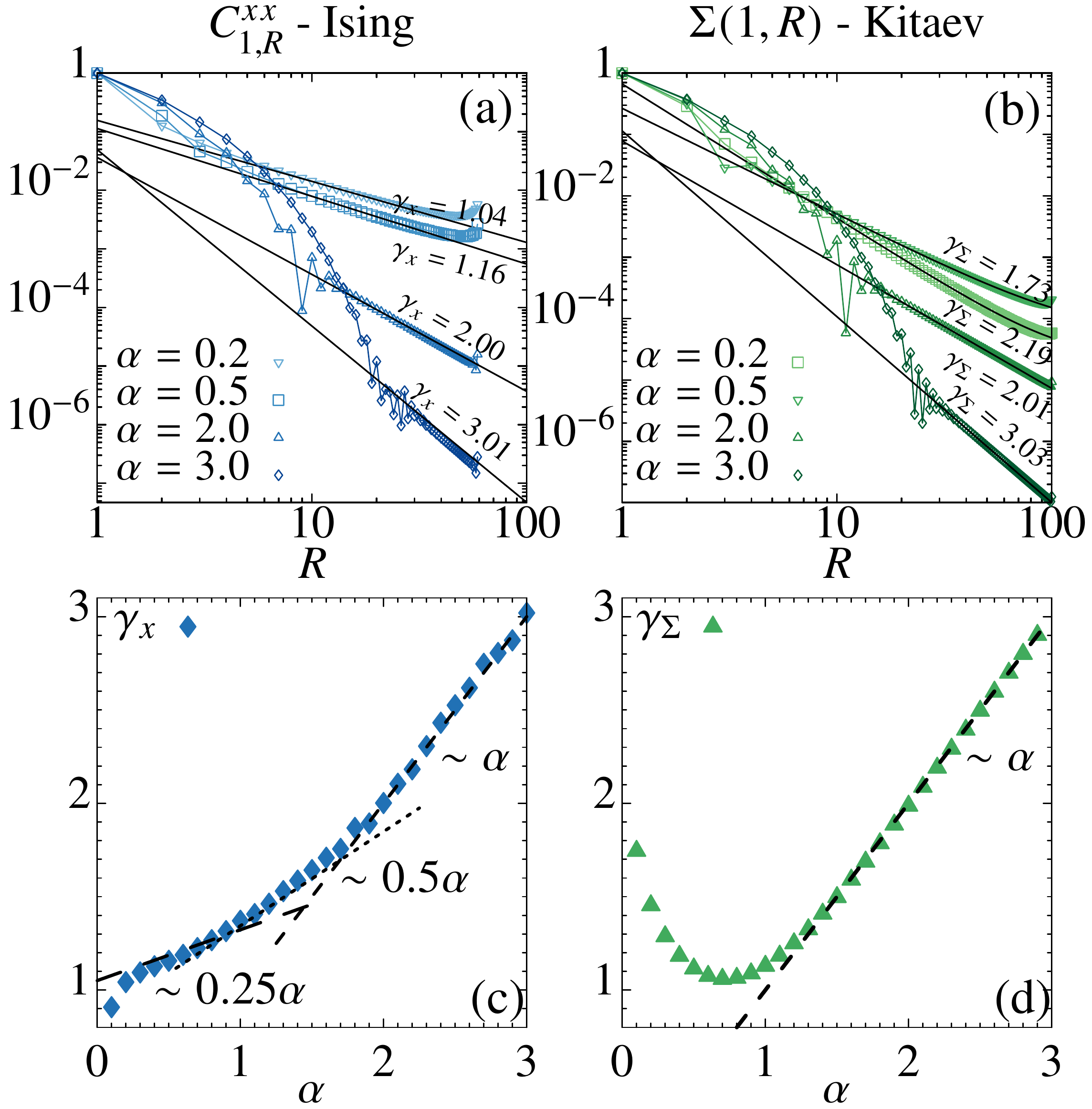}
\caption{(a) $C^{xx}_{1,R}$ correlation Eq.~\eqref{corr:Ising} for $H_\text{Ising}$ ($\theta=0.2\pi$ and  $L=60$), showing the hybrid exponential and power-law behavior for $\alpha \gtrsim 1$ and the purely power-law for $\alpha\lesssim 1$. The oscillations and the bending of $C^{xx}_{1,R}$ for $R\gtrsim 50$ are due to finite size effects of the DMRG simulations.  (b) $\Sigma(1,R)$ correlation Eq.~\eqref{eqn:string} for $H_{\mathrm{LRK}}$ at $\theta=0.2\pi$ and  $L=100$ (PM phase), showing the same hybrid behavior and the same decaying exponent as $C^{xx}_{1,R}$ for $\alpha>1$. For $\alpha<1$ instead the power-law tails have  exponents differing from $C^{xx}_{1,R}$. (c) Decaying exponent $\gamma_x$ of the algebraic tail of the $C^{xx}_{1,R}$ correlation fitted as $1/R^{\gamma_x}$. Three different behaviours for $\gamma_x$, corresponding to the three dashed black lines are numerically observed (see Sec.~\ref{sec:phaseIsing}). (d) Decaying exponent $\gamma_\Sigma$ of the algebraic tail of the $\Sigma(1,R)$ correlation fitted as $1/R^{\gamma_\Sigma}$. }
\label{fig:EEXX}
\end{figure}

We find that for $\alpha\gtrsim1$, $c_{\rm eff}$ is zero everywhere except along two critical lines. By comparing with results for the energy gap (not shown), we find that the critical lines separate two gapped regions [denoted as PM1 and AM in Fig.~\ref{fig1}(a)] that for $\alpha\rightarrow \infty$ correspond to the known paramagnetic and antiferromagnetic phases of the SR model. Similar to Ref.~\cite{Koffel2012}, we find that the behaviour of the {\it full} correlation functions $\braket{\sigma_i^x \sigma_j^x}$ and $\braket{\sigma_i^z \sigma_j^z}$ is consistent with the persistence of paramagnetic and antiferromagnetic orders for all $\alpha$. However, different from the SR model, we find that the  {\it connected} correlation functions decay with distance with a hybrid behaviour that is exponential at short distances and algebraic at long ones. An example is shown in Fig.~\ref{fig:EEXX}(a) for $C^{xx}_{1,R}$ in the PM1 phase. Surprisingly, we find numerically that the exponent $\gamma_x$ of the long-distance decay for $C^{xx}_{1,R}$ displays three difference behaviours: (i) for $\alpha > 2$ it fulfills $\gamma_x=\alpha$, consistent with the results of Refs.~\cite{Deng2005,Koffel2012}. However, (ii) for $1< \alpha < 2$ we obtain a hybrid exponential and algebraic decay with a different $\gamma$ that depends linearly on $\alpha$ with a slope consistent with $\sim 0.55(5)$ and (iii) for $\alpha \lesssim 1$ we observe numerically  a curve compatible with  a pure algebraic decay, with an $\alpha$-dependence of $\gamma_x$ that is linear with slope $\sim 0.25(2)$. The fitted exponent $\gamma_x$ is shown in Fig.~\ref{fig:EEXX}(c). \\

For $\alpha \lesssim 1$ in the paramagnetic regions of the phase diagram denoted as PM2 we find that the effective central charge grows continuously with decreasing $\alpha$ from zero to a finite value that appears to be $\theta$-dependent and has a maximum of order 1 for $\alpha=0$ and $\theta \approx \pi/2$. An example for $\theta = 0.2 \pi$ is plotted in Fig.~\ref{fig1}(d) (blue squares). As mentioned above, in this PM2 region, the correlation function $C^{xx}_{1,R}$ is found numerically to decay as an almost pure power-law.

The energy spectrum changes in this region PM2 compared to the case PM1: the energy gaps $\Delta E(1)$ and $\Delta E(2)$ between the ground state and the first excited states in the odd parity sector increase with decreasing $\alpha$, as shown in Fig.~\ref{DKLRmassive}(b), and the wave-functions of the two lowest-energy excited states $\ket{1^-}$ and $\ket{2^-}$, defined in Sec.~\ref{sec:edgeStates}, become localized at the edges of the chain [Fig.~\ref{edgeDLRK}(b)].\\

In the antiferromagnetic phase, denoted as AM, the effective central charge is instead zero for all $\alpha$~\cite{Koffel2012}. For $\alpha>1$ the connected correlation functions $C^{zz}_{1,R}$ display a clear algebraic decay at long-distances [see the example in Fig.~\ref{fig:ZZIsing}(a) below], while our numerical results do not allow for establishing whether an initial exponential decay is also present, as expected. 
The ground-state is found to be doubly degenerate for all $\alpha$. 
This degeneracy is due to the spontaneous breaking of the $\mathds{Z}_2$ spin-flip symmetry~\cite{diFrancesco1997,Mussardo2010}, and is related to the two modes that are localized at the edges of the chain as in the short-range Ising model~\footnote{The presence of massless edge modes in the AM phase of the LRI model is not a sign of symmetry-protected topological order, as it is discussed for the short-range Ising model in, e.g.,~\cite{Greiter2014}.}.
While a LR power-law tail may be present~\cite{Gong2015}, the localization of these modes is here found to be consistent with exponential up to numerical precision [see Fig.~\ref{edgeDLRK}(b) below]. We come back to this point~below. \\

\subsubsection{LR Kitaev models.}\label{sec:IntroCorrelationKitaev}
The phase diagram of the LRK model Eq.~\eqref{double} for $\epsilon=0$ is reported in Fig.~\ref{fig1}(b), where we plot the effective central charge $c_{\rm eff}$ defined in Sec.~\ref{sec:entropy} as a function of the angle $\theta$ and the power $\alpha$ of the decay of the pairing term. In this case the invariance of Eq.~\eqref{double} under $\cos\theta \to -\cos\theta$ is lost for any finite $\alpha$ and the phase diagram is not symmetric around $\theta=\pi/2$.

Figure~\ref{fig1}(b) shows that for $\alpha\gtrsim 1$ and $0<\theta<\pi$,  two phases exist that are denoted as PM and AM1 separated by two critical lines. In the limit $\alpha \rightarrow \infty $ these phases correspond to the paramagnetic and antiferromagnetic phases of the LRK model and are gapped. Consistently, Fig.~\ref{fig1}(b) shows that the effective central charge $c_{\rm eff}$ is zero within the phases for all $\alpha\gtrsim1$ [see Fig.~\ref{fig1}(d) for an example], while $c_{\rm eff}\neq 0$ along the critical lines, as expected from general results for SR systems~\cite{Calabrese2004}. \\

The PM and AM1 phases are distinguished by different asymptotic values of the correlation functions $\Sigma(i,j)$  defined in Sec.~\ref{sec:Correlation}. In the region denoted as AM1,  $\Sigma(i,j)$  has a finite value for $\abs{i-j}\to\infty$, while $\Sigma(i,j)$ decays  for $\abs{i-j}\to\infty$ within the PM phase. Similar to the situation in the LRI model (see above),  the decay to zero of $\Sigma(i,j)$ in the PM phase shows a hybrid exponential and power-law behaviour with distance. This is shown for a particular value of $\theta$ in Fig.~\ref{fig:EEXX}(b), where we find numerically that the exponent $\gamma_\Sigma$ for the power-law tail of $\Sigma(i,j)$ equals $\gamma_z=\alpha$ when $\alpha>1$. For $\alpha<1$, however, the exponential part becomes numerically unobservable, and $\Sigma(i,j)$ decays essentially algebraically within the PM phase with an exponent that grows to 2 for $\alpha\rightarrow 0$ [see Fig.~\ref{fig:EEXX}(b)].\\

Remarkably, we show below in Sec.~\ref{sec:CorrelationAnalytical} that the hybrid exponential and algebraic behaviour described above can be obtained analytically in all phases for several correlation functions, such as the one-body and the density-density correlation functions. In particular, the leading contribution to the one-body correlation function $\braket{a^\dag_i a_j}$ reads 
\begin{equation}
\braket{a^\dag_R \, a_0} =\mathcal{A}_{\alpha,\theta} \cdot \frac{(-1)^R \nepero^{-\xi R}}{\sqrt{R}} +  \mathcal{B}_{\alpha,\theta} \cdot \begin{cases}
\dfrac{1}{R^{\alpha+1}} & \alpha>2 \\[1em]
\dfrac{1}{R^{2\alpha-1}} & 1<\alpha<2 \\[1em]
\dfrac{1}{R^{2-\alpha}} & 0<\alpha<1 
\end{cases}
\label{Green}
\end{equation}
where the pre-factors $\mathcal{A}_{\alpha,\theta}$ and $\mathcal{B}_{\alpha,\theta}$ are derived below in Sec.~\ref{sec:CorrelationAnalytical}. The  algebraic part of the decay of $g_2(R)$ is instead found to be $g_2(R)\sim R^{-2\alpha}$ and $g_2(R)\sim R^{-2}$ for $\alpha > 1$ and $\alpha<1$, respectively.\\ 

In Sec.~\ref{sec:MajoranaModes}, we show that a similar hybrid exponential and power-law decay is found for the localization of the edge modes within the antiferromagnetic phase AM1: For $\alpha \gtrsim 1$, the edge modes are massless, as expected from the SR Kitaev model.  However, for $\alpha \lesssim 1$ the edge modes acquire a {\it finite mass}, i.e., become gapped. This is shown in Fig.~\ref{fig1}(c), where we plot the edge gap $\Lambda_{0}$ defined in Sec~\ref{sec:edgeStates} as a function of $\alpha$ for a few values of the parameter $\epsilon$ of Hamiltonian Eq.~\eqref{double}: while for $\alpha \gtrsim 1$ the gap scales to zero with the system size $L$ as $1/L^\alpha$, for $\alpha\lesssim 1$ it remains finite and for $\alpha\sim0$ can be of order unity. The presence of the gap removes the degeneracy of the ground-state, signaling a new phase for this class of topological LR models. This latter phase is denoted as AM2 in Fig.~\ref{fig1}.

\subsection{Critical lines}\label{sec:HamD}\label{sec:CriticalLines}

\subsubsection{LR Kitaev models.}
For the LRK models, the critical lines can be computed exactly as Eq.~\eqref{double} are integrable.  A Fourier transform of the fermionic operators $a_j$ takes Eq.~\eqref{double} to the form
\begin{equation}\label{eqn:KitaevMomentum}
H_\text{LRK}= \sin\theta \sum_k \begin{pmatrix}a^\dag_k & a_{-k}\end{pmatrix} \begin{pmatrix} 
g_\alpha(k) +\cot\theta & -\uImm (1+\epsilon)f_\alpha(k) \\ 
\uImm  (1+\epsilon) f_\alpha(k) & -(g_\alpha(k) +\cot\theta ) \\ 
\end{pmatrix}
\begin{pmatrix}a_k \\ a^{\dagger}_{-k}\end{pmatrix} 
\end{equation}
where $a_k$ reads $a_k=\frac{1}{\sqrt{L}} \sum_j \nepero^{\uImm k j} a_j$, $k=2\pi n/L$ is the lattice momentum with $n=0,\dots,L-1$, and the functions $f_\alpha(k)$ and $g_\alpha(k)$ read $f_\alpha(k)=\sum_\ell \sin(k \ell)/\ell^\alpha$ and  $g_\alpha(k)=\sum_\ell \cos(k \ell)/\ell^\alpha$, respectively. \\

A Bogoliubov transformation brings Eq.~\eqref{eqn:KitaevMomentum} in diagonal form as
\begin{equation}\label{eqn:LRKmomentum}
H_{\mathrm{LRK}}= \sum_{k} \lambda_\alpha(k) \left(\eta^\dag_k\eta_k  - 1/2\right),
\end{equation}
with 
\begin{equation}\label{eqn:dispersion}
\lambda_\alpha(k)=2\sin\theta \sqrt{(g_\alpha(k) +\cot \theta)^2 + (1+\epsilon)^2f_\alpha(k)^2}.
\end{equation} 
Here, the new fermionic operators $(\eta^\dag_k,\eta_{-k})$  are given in terms of the operators $(a^\dag_k,a_{-k})$ by
\begin{equation}
\left(\begin{array}{c}\eta^\dag_k \\ \eta_{-k}\end{array}\right) = \left(\begin{array}{cc} \sin \beta_k  &  \uImm \cos\beta_k \\ \uImm \cos\beta_k  & \sin\beta_k  \end{array}\right) \left(\begin{array}{c} a^\dag_k \\ a_{-k}\end{array}\right),
\end{equation}
where $\tan (2\beta_k) = (1+\epsilon)f(k)/[g(k)+\cot\theta]$. The ground state of Eq.~\eqref{eqn:LRKmomentum} is the vacuum of the $\eta_k$ fermions and has an energy density $e_0(\alpha,L) = - \sum_k \lambda_\alpha(k)/(2L)$. \\

The critical lines can be computed from the dispersion relation Eq.~\eqref{eqn:dispersion} as follows: (i) For a finite system with $L$ sites, Eq.~\eqref{eqn:dispersion} is zero on the line $\cot \theta + g_\alpha(0)=0$ and  on the line $\cot \theta + g_\alpha(\pi)=0$; (ii) For a system in the thermodynamic limit $L\to\infty$ instead  \(f_\alpha(k)=\Im \Li{\alpha} \, \big(\nepero^{\uImm k}\big)\) and \(g_\alpha(k)=\Re \Li{\alpha} \, \big(\nepero^{\uImm k}\big)\), $\Li{\alpha}(x)$ being the polylogarithm of order $\alpha$~\cite{Frank2010},  and the critical line $\cot \theta + g_\alpha(0)=0$  ends at the point $\theta=\pi$ and $\alpha=1$.\\

For the critical line with $\theta<\pi/2$ we compute the value of the central charge $c$ by two methods: (i) by fitting the von Neumann entropy Eq.~\eqref{vN} and (ii) by studying the finite size corrections to the ground state energy density $e_0(\alpha,L)$ (see Ref.~\cite{Vodola2014} for a similar model). 

The results for $c$ obtained from the scaling of the von Neumann entropy Eq.~\eqref{vN} are reported in Fig.~\ref{fig:crossingMethod}(b). For $\alpha \gtrsim 1$, we find $c=1/2$ as expected from the SR model. For $\alpha \lesssim 1$, however, $c_{\mathrm{eff}}$ increases up to values of order one [see red dashed line in Fig.~\ref{fig1}(a)]. In Ref.~\cite{Vodola2014} it was demonstrated for a related model with LR pairing only that this behaviour corresponds to an exotic change for the decay of density-density correlation functions: For $\alpha \lesssim 1$ their oscillations mimic those of a Luttinger liquid. Here, we find a similar behaviour (not shown). The increasing of $c_{\mathrm{eff}}$, below $\alpha=1$, is also found in the very recent work~\cite{Ares2015} where the scaling of the von Neumann entropy in the thermodynamic limit is analytically analized.

This anomalous behaviour of $c_{\mathrm{eff}}$ points towards a breaking of conformal symmetry along the critical line, which we analyze further below.\\

The breaking of conformal symmetry can be inferred also by analyzing the scaling of the energy density $e_0(\alpha, L)$  with the system size $L$~\cite{Vodola2014}:
For a conformally invariant theory the following relation must hold~\cite{Henkel1999}
\beq
e_{0}(\alpha, L) = e_{\infty}(\alpha) - \frac{\pi \, v_F \, c}{6 L^2} \,,
\label{scalinge}
\eeq 
where $e_{\infty}(\alpha)$ is the energy density in the thermodynamic limit, $v_F=\abs{2 \sin \theta \, \mathrm{Li}_{\alpha-1}(-1)}$ is the Fermi velocity and $c$ is the central charge of the conformal theory.
We analyzed numerically $e_{0}(\alpha, L) $, finding that relation \eqref{scalinge} works properly only for sufficiently large $\alpha \gtrsim 2$. This results in a value $c \simeq 1/2$ for $\alpha \gtrsim 2$, as expected from results for the SR Kitaev chain. Conversely, for $\alpha \lesssim 2$, $e_{0}(\alpha, L)$ does not satisfy the scaling law \eqref{scalinge}, which implies a breaking of conformal symmetry.
This behaviour also implies that the quantum phase transition  between the PM and the AM1 phase for $\theta<\pi/2$ and $\alpha \lesssim 2$ is in a different universality class from that of the SR Kitaev (Ising) model.
We notice that, even if conformal symmetry is broken, the von Neumann entropy predicts a value  for $c_\text{eff}$ which tends to $1$ as $\alpha$ goes to zero, compatible with the observed decay of density-density correlation functions. \\

We further confirm the breaking of conformal symmetry for the fermionic models by looking at the behaviour of their low-energy spectra. The dispersion relation of a conformal field theory is linear in the momentum $k$, implying a dynamical exponent $z=1$~\cite{Henkel1999}. Consistently, by expanding the dispersion relation $\lambda_\alpha(k)$ for the long range Kitaev models on the critical line with $\theta>\pi/2$,  for $\alpha>2$ we find $\lambda_\alpha(k) \sim k$, for $k \to 0$. However,  for $1<\alpha<2$ we obtain the scaling $\lambda_\alpha(k) \sim k^{\alpha-1}$. This latter scaling implies a dynamical exponent $z=\alpha-1$ that varies continuously  with $\alpha$ and is different from that of a conformal field theory.  This would imply that the quantum phase transition  between the PM and the AM1 phase for $\theta>\pi/2$ and $\alpha<2$ is in a different universality class from that of the short-range Kitaev model. The appearance of a new universality class due to long-range interactions is also found in Refs.~\cite{Maghrebi2015_2, Gong2015_2}. Incidentally, we notice that linearity of the spectrum around the minimum is only a necessary condition for the persistence of conformal invariance: indeed along  the critical line at $\theta<\pi/2$ conformal symmetry breaking arises even if the low-energy spectrum is linear for every $\alpha$.

\subsubsection{LR Ising model.}

\begin{figure}\centering
\includegraphics[width=0.5\textwidth-4pt]{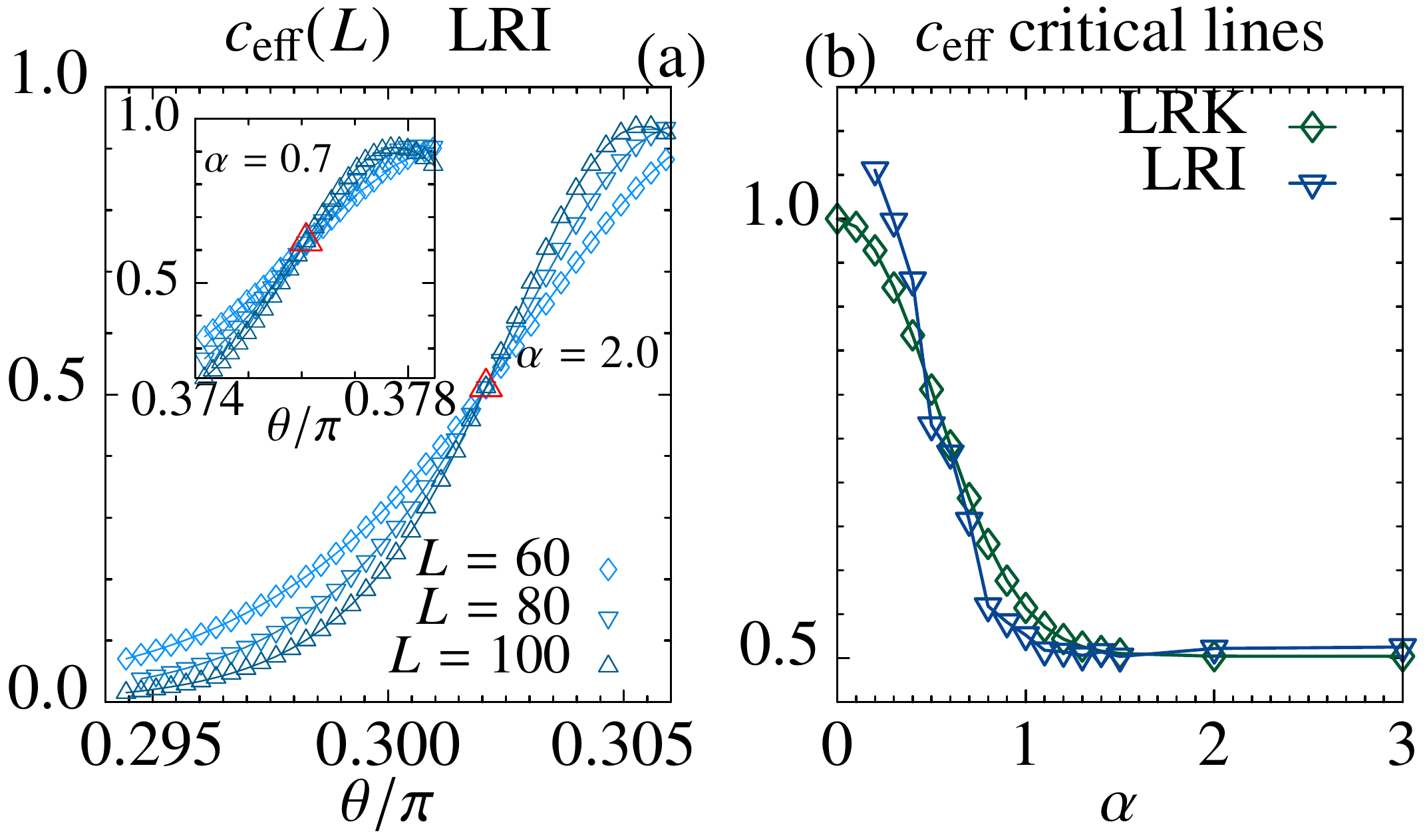}
\caption{(a) Effective central charge extracted from the scaling with $L$ of the von Neumann entropy \eqref{vN} as function of $\theta$ for two values of $\alpha$, plotted for different system sizes $L=60,80,100$. We located the critical point (indicated by red triangles)  where the effective central charge does not depend on $L$. (Main panel) scaling at $\alpha=2.0$: the critical point is $\theta_c\approx  0.302\pi$ and  the central charge $c$ is compatible with 1/2. (Inset) scaling at $\alpha=0.7$: the critical point is $\theta_c\approx 0.376\pi$ and the central charge $c$ is compatible with $c\approx 0.66$. (b) Central charge on the critical lines of the LRI model (blue triangles) obtained with the same method as in panel (a) and of the LRK model (green diamonds) obtained  by finite size scaling from the expression for entanglement entropy in Eq.~\eqref{vN}.}\label{fig:crossingMethod}
\end{figure}

We locate the critical line (for $\theta<\pi/2$, the other being symmetric) of the LRI model numerically by using two complementary ways that agree up to finite-size effects. Firstly, we determine the points in the phase diagram of Fig.~\ref{fig1}~(a) where the energy gap between the ground state and the first excited state reaches its minimum. Then, we compute the effective central charge $c_{\rm eff}$ for different system sizes $L$ from Eq.~\eqref{vN} and determine the precise values of $\alpha$ and $\theta$ for which its value does not depend on $L$~\cite{CamposVenuti2006,Roncaglia2008}. Examples of this latter technique, which is found to be particular precise, are presented in Fig.~\ref{fig:crossingMethod} for different $\alpha$. We notice however that this method does not allow us to extract a precise value for $c_{\mathrm{eff}}$ when $\alpha \lesssim 0.2$, since within our numerical results lines with different $L$ do not cross at a single point in this region.\\

On the critical line, we find that for $\alpha \gtrsim 1$  (black solid lines in Fig.~\ref{fig1}) $c_{\rm eff}$ is equal to $1/2$ as expected for  the central charge of the critical SR Ising model. However, for $\alpha \lesssim 1$ (red dashed lines in Fig.~\ref{fig1}), $c$ increases continuously up to a value of order 1 as shown in Fig.~\ref{fig:crossingMethod}(b). We argue that on this line the conformal invariance of the model is broken as the found values of $c$ do not coincide with the discrete set allowed for the known conformal field theories~\cite{diFrancesco1997, Mussardo2010}. 

Based on the mismatch between the predictions for $c$ from the von Neumann entropy and the ground state energy density found in the previous subsection for the LRK model, we cannot exclude here a conformal symmetry breaking also in a certain range for $\alpha$ above $\alpha =1$. However our results do not allow us to provide a final answer, since our DMRG calculations for the LRI model cannot reach sufficiently large sizes to perform a satisfying finite-size scaling for the energy density.

\section{Correlation functions for the LRK}\label{sec:Correlations}\label{sec:CorrelationAnalytical}

\begin{figure}\centering
\includegraphics[width=0.5\textwidth-4pt]{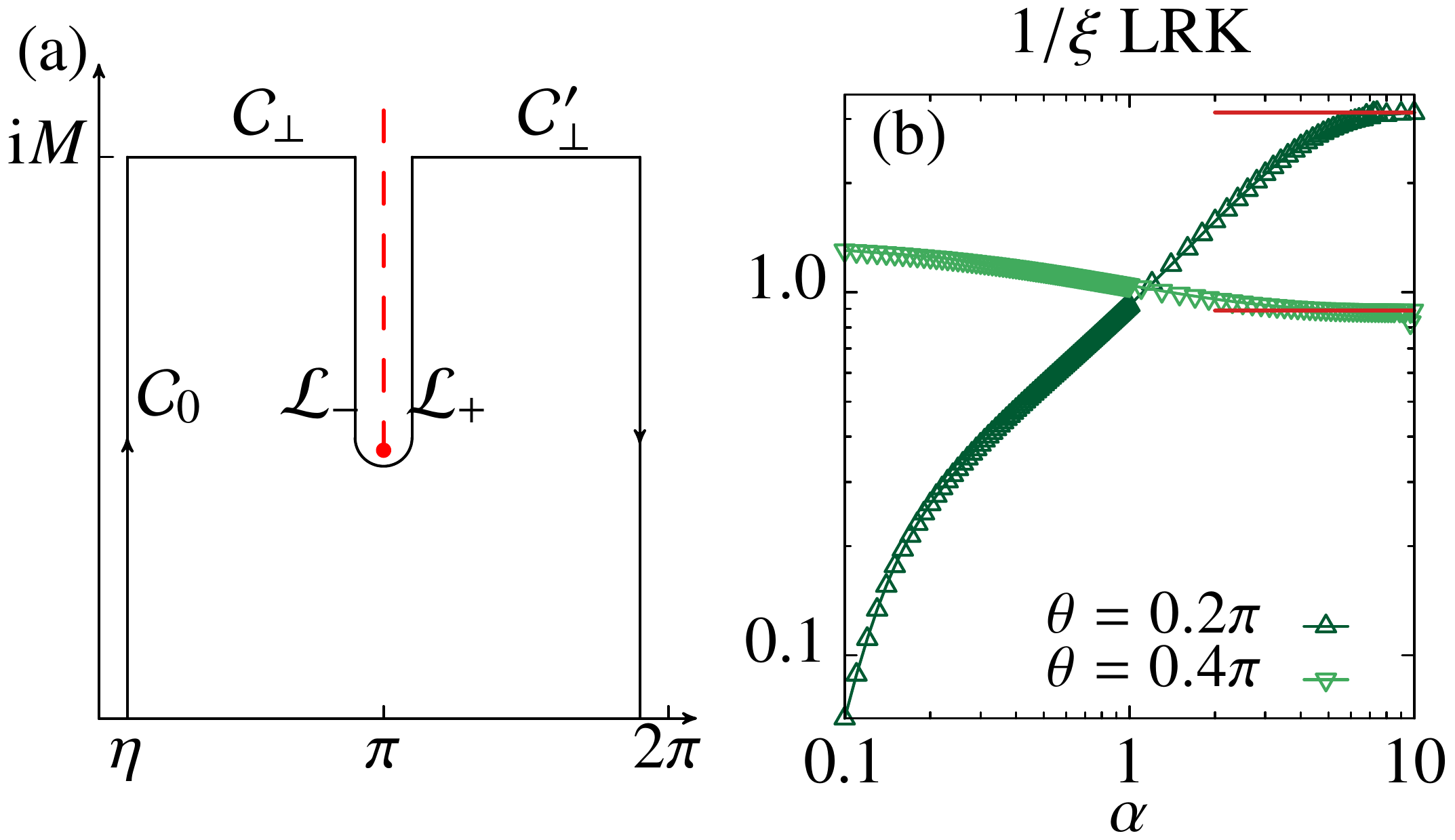}
\caption{(a) Integration contour adopted to evaluate the integral \eqref{eqn:NormalCorrelatorContinuum}. The red dashed line is the branch cut of the square root in the denominator of the integrand in \eqref{eqn:NormalCorrelatorContinuum}. (b) Decay rate $1/\xi$ of the exponential part of the correlation of Eq.~\eqref{eqn:CorrExp} as function of $\alpha$ for two values of $\theta$. The red lines represent the value of $1/\xi$ for $\alpha\to\infty$.}\label{fig:contour}
\end{figure}

In this Section, we present an analytical calculation of the one-body correlation functions for the LRK models. The latter display a hybrid exponential and algebraic decay with distance that is explained by exploiting the integrability of the models. Higher-order correlation functions, such as the density-density correlations, are readily obtained from these correlations via Wick's theorem [see Sec.~\ref{sec:IntroCorrelationKitaev} and below]. \\
 
The one-body correlation functions $\braket{a^\dag_R a_0}$ and $\braket{a^\dag_R a^\dag_0}$ read
\begin{equation}
\braket{a^\dag_R a_0}  = - \frac{\sin\theta}{2\pi}  \int_{0}^{2\pi}\de k \,\nepero^{\uImm k R }\left(  \frac{ g_\alpha (k) + \cot \theta}{2\lambda_\alpha(k)}\right) \label{eqn:NormalCorrelatorContinuum} 
\end{equation}
and 
\begin{equation}
\braket{a^\dag_R a^\dag_0}  = \frac{\uImm \sin\theta}{2\pi}  \int_{0}^{2\pi}\de k \,\nepero^{\uImm k R }\left(  \frac{ f_\alpha (k) }{2\lambda_\alpha(k)}\right) \, ,\label{eqn:AnomalousCorrelatorContinuum} 
\end{equation}
respectively. In the following, we focus first on the one-body correlation function $\braket{a^\dag_R a_0}$ and come back to the anomalous correlation $\braket{a^\dag_R a^\dag_0}$  in Sec.~\ref{HybExpAl} below.\\

In order to evaluate Eq.~\eqref{eqn:NormalCorrelatorContinuum}, we use the Cauchy theorem applied to the contour in the complex plane drawn in Fig.~\ref{fig:contour}
\begin{equation}
\braket{a^\dag_R a_0}  = -\frac{1}{2\pi} \left( \int_{\mathcal{C}_0}  + \int_{\mathcal{L}_{-}} + \int_{\mathcal{L}_{+}} + \int_{\mathcal{C}_{2\pi}}  \right) \, \nepero^{\uImm z R}\ \mathcal{G}_\alpha(z) \, \de z
\label{int}
\end{equation}
with $\mathcal{G}_\alpha(z)=[g_\alpha(z)+\cot\theta]/[2 \sqrt{(\cot\theta + g_\alpha(z))^2 + f^2_\alpha(z)}]$ and $z=k+\uImm y$, and where we have chosen $\epsilon=0$. In Eq.~\eqref{int} we have neglected the contributions from \(\mathcal{C}_\perp\) and \(\mathcal{C}'_\perp\) as they vanish when \(M\to \infty\). As we explain below, the integrations over the lines $\mathcal{L}_{\pm}$ and $\mathcal{C}_{0,2\pi}$ (and thus momenta $k\simeq \pi$ and $k\simeq 0$ and $2\pi$, respectively) are responsible for the exponential and algebraic behaviour observed in these models, respectively. 

\subsection{Exponential decay}\label{sec:CorrA}
The sum of the integrals on the lines $\mathcal{L}_{-}$ (where $z=\pi^-+\uImm y$) and $\mathcal{L}_{+}$ (where $z=\pi^++\uImm y$) of Fig.~\ref{fig:contour} gives
\begin{equation}\label{eqn:ExponentialPart}
I^\text{exp}_{\alpha}(R) = - \frac{ \nepero^{\uImm \pi R} \,  \nepero^{-  \xi R}}{\pi}  \int_{0}^{\infty} \, \nepero^{- y R} \, \mathcal{G}_\alpha(\pi +\uImm (y+\xi)) \, \de y .
\end{equation}
Equation~\eqref{eqn:ExponentialPart} displays an exponential behaviour with a decay constant $1/\xi$. The appearance of this quantity is due to the square root in the denominator of  $\mathcal{G}_\alpha(z)$, yielding a branch cut from $z_1=\pi+\uImm \, \xi$ to $\infty$. The leading term in Eq.~\eqref{eqn:ExponentialPart} is obtained by integrating $\mathcal{G}_\alpha(\pi +\uImm (y+\xi))$ in the limit  \(y\to 0\) \cite{Ablowitz2003}, and reads
\begin{equation}
I^\text{exp}_{\alpha}(R)=  (-1)^R \mathcal{A}_{\alpha,\theta} \frac{ \nepero^{-\xi R}}{\sqrt{R}},
\label{eqn:CorrExp}
\end{equation}
with
\begin{equation}\label{eqn:constant1}
\mathcal{A}_{\alpha,\theta} =  -\frac{(2\cot\theta+\Li{\alpha}(-\nepero^{\xi})+\Li{\alpha}(-\nepero^{-\xi}))}{4\sqrt{\pi}\abs{\Li{\alpha-1}(-\nepero^{\xi})}^{1/2}[\cot\theta+\Li{\alpha}(-\nepero^{-\xi})]^{1/2}} \, .
\end{equation}

The decay constant $\xi$ is related to the zeroes of the denominator of $\mathcal{G}_\alpha(z)$ and is obtained by solving the equation 
\begin{equation}\label{eqn:ZeroDispersionRelation}
[\cot\theta + \Li{\alpha}(-\nepero^{-\xi})][\cot\theta + \Li{\alpha}(-\nepero^{\xi})]=0.
\end{equation}

Two cases must be distinguished: if $\cot\theta<\abs{\Li{\alpha}(-1)}$, the equation $[\cot\theta + \Li{\alpha}(-\nepero^{-\xi})] = 0$ admits a solution, since the function $-\Li{\alpha}(-\nepero^{-\xi})$ is always decreasing for $\xi>0$. If instead $\cot\theta>\abs{\Li{\alpha}(-1)}$, then $[\cot\theta + \Li{\alpha}(-\nepero^{\xi})] = 0$ admits a solution, for the same reason as above.
In the following we focus, without loss of generality, on the first case, where $\xi$ is solution of $\cot\theta=-\Li{\alpha}(-\nepero^{-\xi})$.\\
Notably for $\alpha=0$ solutions exist only for $\theta>\pi/4$ [in which case, $\xi=\pm\log(\tan\theta-1)$ for $\cot\theta \lessgtr 1/2$]. For $\theta<\pi/4$, instead,  Eq.~\eqref{eqn:ZeroDispersionRelation} does not admit any solutions, which
implies the {\it absence} of exponential decay. This is in contrast with, e.g., the expected behaviour of correlation functions within gapped phases for SR models.\\

Figure~\ref{fig:contour}(b) shows the decay constant $1/\xi$ as a function of $\alpha$ for two different values of $\theta$. In particular, for $\theta=0.2\pi$ and $\alpha\to\infty$, $1/\xi$ tends to the SR value ($\xi\to\abs{\log\abs{\tan\theta}}$), as expected. However, for $\alpha \to 0$ we find that $1/\xi$ tends to zero, essentially linearly with $\alpha$. As explained below, this can result in the non observability of the exponential dependence of correlation functions for $\alpha\lesssim 1$. Notice however that even if $\xi$ is finite when $\alpha\lesssim 1$ for $\theta=0.4\pi$, the exponential part of the correlation functions is still unobservable.

\subsection{Algebraic decay}\label{sec:CorrB}

The sum of the integrals on the lines $\mathcal{C}_0$ (where $z=\eta + \uImm y$) and $\mathcal{C}_{2\pi}$ (where $z=2\pi - \eta + \uImm y$) of Fig.~\ref{fig:contour} gives
\begin{equation}
I^\text{pow}_\alpha(R)= \frac{1}{\pi}  \int_0^\infty   \,\nepero^{- y R} \,\Im  \mathcal{G}_\alpha(\uImm y) \, \de y,
\label{eqn:IntegralZeroTwoPi}
\end{equation}
after sending $\eta \to 0$. The leading contribution for $R\to\infty$ to the integral in Eq.~\eqref{eqn:IntegralZeroTwoPi}  can be computed again by integrating the imaginary part $\Im \mathcal{G}_\alpha(\uImm y)$ for $y\to 0$. By exploiting the following series expansion of the polylogarithm~\cite{Frank2010}
\begin{equation}
\Li{\alpha}(\nepero^{\pm y}) = \Gamma(1-\alpha)(\mp y)^{\alpha-1}+ \sum_{j=0}^\infty \frac{\zeta(\alpha-j)}{j!}(\pm y)^j,
\end{equation}
one obtains
\begin{equation}
I^\text{pow}_\alpha(R)= \mathcal{B}_{\alpha,\theta}\cdot
\begin{cases}
\dfrac{1}{R^{\alpha+1}} & \alpha > 2 \\[1em]
\dfrac{1}{R^{2\alpha-1}} & 1<\alpha<2\\[1em]
\dfrac{1}{R^{2-\alpha}} & \alpha<1
\end{cases}\label{eqn:CorrPowerLaw}
\end{equation}
with 
\begin{equation}
\mathcal{B}_{\alpha,\theta}= 
\begin{cases}
\dfrac{ \alpha\zeta(\alpha-1) }{(\zeta(\alpha)+\cot\theta)^2}  & \alpha > 2 \\[1em]
\dfrac{(\cos (\pi  \alpha )-1) \Gamma (1-\alpha ) \Gamma (2 \alpha -1)}{4 \Gamma (\alpha ) (\cot (\theta )+\zeta (\alpha ))^2}  & 1<\alpha<2 \\[1em]
\dfrac{\cos ^3\left(\frac{\pi  \alpha }{2}\right) \Gamma (2-\alpha ) (\cot (\theta )+\zeta (\alpha ))}{\pi \, \Gamma (1-\alpha ) } & \alpha<1 
\end{cases}\label{eqn:PreB}
\end{equation}
While, e.g., the phase diagram of Fig.~\ref{fig1}(b) demonstrates the persistence of individual paramagnetic and antiferromagnetic phases with varying $\alpha$, the analytic expressions Eqs.~\eqref{eqn:CorrPowerLaw} and~\eqref{eqn:PreB} for the one-body correlation function clearly show that different regions are in fact present within each phase.\\

\subsection{Hybrid decay and other correlations}\label{HybExpAl}
\begin{figure}\centering
\includegraphics[width=0.5\textwidth-4pt]{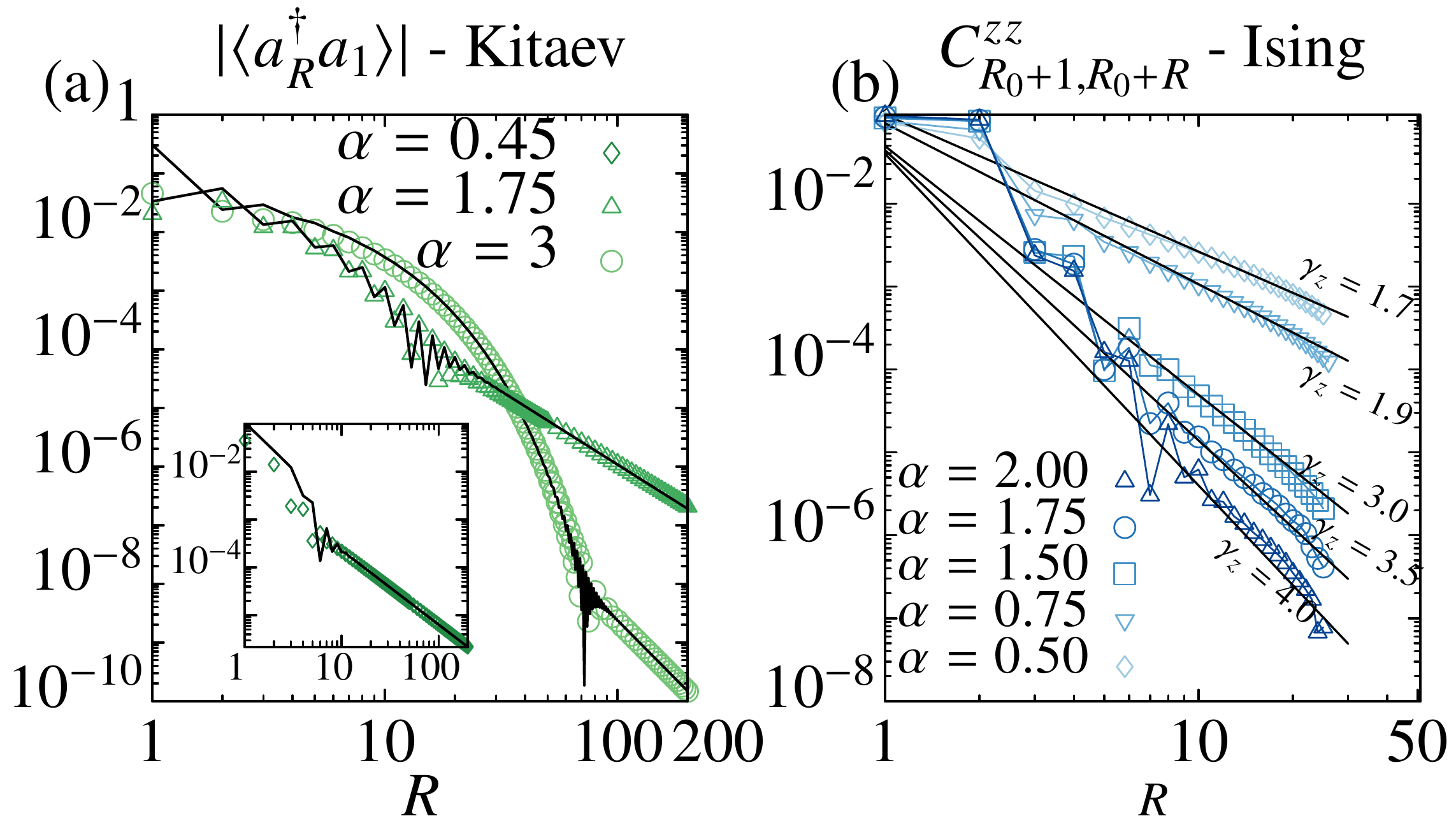}
\caption{ (a) Comparison between numerical solutions of Eq.~\eqref{eqn:NormalCorrelatorContinuum} (symbols) and the analytic expressions Eqs.~\eqref{eqn:CorrExp} and \eqref{eqn:CorrPowerLaw} (solid lines) for $\theta=0.239\pi$ and different $\alpha$.  The inset refers to $\alpha=0.45$, where the exponential part is negligible. (b)  $C^{zz}_{R_0+1,R_0+R}$ correlation for $H_\text{LRI}$  at $\theta=0.207\pi$, $L=100$, $R_0=L/4$ and different $\alpha$. The power-law part of $C^{zz}_{R_0+1,R_0+R}$ shows the same decay  exponent $\gamma_z=2\alpha$ for $\alpha >1$, as $g_2(0,R)$  for $H_{\mathrm{LRK}}$ (Eq.~\eqref{eqn:DensDensKitaev}, solid black lines).}
\label{fig:CorrelationsAnalytical}
\end{figure}

The two contributions from Eqs.~\eqref{eqn:CorrExp} and \eqref{eqn:CorrPowerLaw} sum up to give the hybrid exponential-algebraic behaviour of Eq.~\eqref{Green} valid for all $\alpha$. Figure~\ref{fig:CorrelationsAnalytical}(a) shows that the analytical results are in perfect agreement with a numerical solution of Eq.~\eqref{eqn:NormalCorrelatorContinuum}.\\

The anomalous correlation $\braket{a^\dag_R a^\dag_0}$ can be computed along the same lines as before and is given by
\begin{equation}\label{eqn:finalAnomalous}
\braket{a^\dag_R a^\dag_0} = \mathcal{A}^{(a)}_{\alpha,\theta}\frac{(-1)^R \nepero^{-\xi R}}{\sqrt{R}} + \mathcal{B}^{(a)}_{\alpha,\theta}\cdot \begin{cases}
 \displaystyle \frac{1}{R^{\alpha}} & \alpha > 1 \\[1.25em]
 \dfrac{1}{R} & 0<\alpha < 1 
\end{cases}
\end{equation}
with
\begin{equation}
\mathcal{A}^{(a)}_{\alpha,\theta}=\frac{1}{4\sqrt{\pi}}  \frac{\Li{\alpha}(-\nepero^{-\xi})-\Li{\alpha}(-\nepero^{\xi})}{\abs{\mathrm{Li}_{\alpha-1}(-\nepero^{\xi})}^{1/2}(\cot\theta + \Li{\alpha}(-\nepero^{-\xi}))^{1/2}} \label{eqn:constant3}
\end{equation}
and
\begin{equation}
\mathcal{B}^{(a)}_{\alpha,\theta}=\begin{cases}
-\dfrac{1}{4  ( \cot \theta +\zeta (\alpha )) }   & \alpha>1 \\[1em]
-\dfrac{1}{2\pi}\cos\dfrac{\pi \alpha}{2} & 0<\alpha<1 \, .
\end{cases}\label{eqn:constant4}
\end{equation}

From $\braket{a^\dag_R a_0}$ and $\braket{a^\dag_R a^\dag_0}$ we can compute  other local correlations  by Wick's theorem. For instance,  the density-density correlation function $g_2(i,j)$ reads $g_2(i,j) = \abs{\braket{a^\dag_i a^\dag_j}}^2 - \abs{\braket{a^\dag_i a_j}}^2$ and, from Eqs.~\eqref{eqn:CorrPowerLaw} and \eqref{eqn:finalAnomalous}, the leading part of its LR power law tail is found to be
\begin{equation}\label{eqn:DensDensKitaev}
g_2(0,R) \sim \begin{cases}
\dfrac{1}{R^{2\alpha}} & \alpha>1 \\[1em]
\dfrac{1}{R^{2}} &  0<\alpha<1.
\end{cases}
\end{equation} 

We notice that a hybrid exponential and algebraic behavior similar to that described above has been already observed numerically in certain spin and fermionic models, e.g., in Refs.~\cite{Deng2005, Koffel2012, Vodola2014, Gong2015}. This behaviour is characteristic of the non-local interactions and from our analysis appears to be largely unrelated to the presence or absence of a gap in the spectrum. 
In fact, we have shown here for Hamiltonians Eq.~\eqref{double} that this hybrid decay does not require exotic properties of the spectrum (see~Refs.~\cite{Gong2014} and~\cite{Foss-Feig2015}), rather is due to the different contributions of momenta $k=\pi$ (as for the SR limit) and  $k = 0, 2\pi$ respectively. In particular, the latter momenta are responsible for the LR algebraic decay. \\

Finally, as mentioned before, we notice that (i) the contribution to the imaginary part in Eq.~\eqref{eqn:IntegralZeroTwoPi} is due to \(\Li{\alpha}(\nepero^{y})\) and disappears in the limit $\alpha\to\infty$. This implies a simple exponential decay of correlations as expected for a SR model. Conversely, (ii) the exponential decay is negligible when $\alpha \lesssim 1$, as shown in the inset of Fig.~\ref{fig:CorrelationsAnalytical}(a), implying an essentially pure algebraic decay.

\subsection{Comparison with correlations of the LRI model}\label{sec:CorrC}
In this section we compare the decay with distance of the correlation functions $g_2(i,j)$  and $\Sigma(i,j)$ of the LRK models with those of the correlations $C^{zz}_{i,j}$ and $C^{xx}_{i,j}$  of the LRI model, respectively, since they are related by a Jordan-Wigner transformation [see, e.g., Sec.~\ref{sec:Correlation}].
The correlation functions $C^{zz}_{i,j}$ and $C^{xx}_{i,j}$  have been studied within the PM phase in Ref.~\cite{Koffel2012} for $\alpha\gtrsim 0.5$. There,  for $\alpha \gtrsim 1$ it was found that the long-distance behaviour is characterized by  an algebraic decay with exponents $\gamma_z = 2 \alpha$ and $\gamma_x = \alpha$ for the two correlations, respectively. \\

For $\alpha\gtrsim 1$, our own calculations for $C^{xx}_{i,j}$ and $C^{zz}_{i,j}$ are reported in Fig.~\ref{fig:EEXX}(b) and Fig.~\ref{fig:CorrelationsAnalytical}(b), respectively. There, we compare the decay of correlations with that of the corresponding correlation functions in the LRK models, $g_2(i,j)$ [see also Eq.~\eqref{eqn:DensDensKitaev}] and $\Sigma(i,j)$, respectively, showing very good quantitative agreement with the (semi-)analytic results. 
In agreement with Refs.~\cite{Deng2005, Vodola2014}, an hybrid exponential and power-law behaviour is found for $C^{xx}_{i,j}$ [as well as for $\Sigma(i,j)$ and $g_2(i,j)$].

Conversely, when $\alpha \lesssim 1$ only an algebraic tail is visible for the decay of $C^{zz}_{i,j}$ in Fig.~\ref{fig:CorrelationsAnalytical} and  $C^{xx}_{i,j}$ [as well as $\Sigma(i,j)$] in Fig.~\ref{fig:EEXX}, since the initial exponential decay is too small to be observed, as expected from the discussion above. Moreover, no universal behaviour for the decay exponents is identifiable for $C^{xx}_{i,j}$ and $C^{zz}_{i,j}$ in this region.  The exponents for $C^{xx}_{i,j}$ and $\Sigma(i,j)$ differ here in general, probably  because the contribution of the string operators in $H_\text{LRI}$ becomes more relevant.\\

\begin{figure}[b]
\centering\hspace{-1em}
\includegraphics[width=0.5\textwidth-4pt]{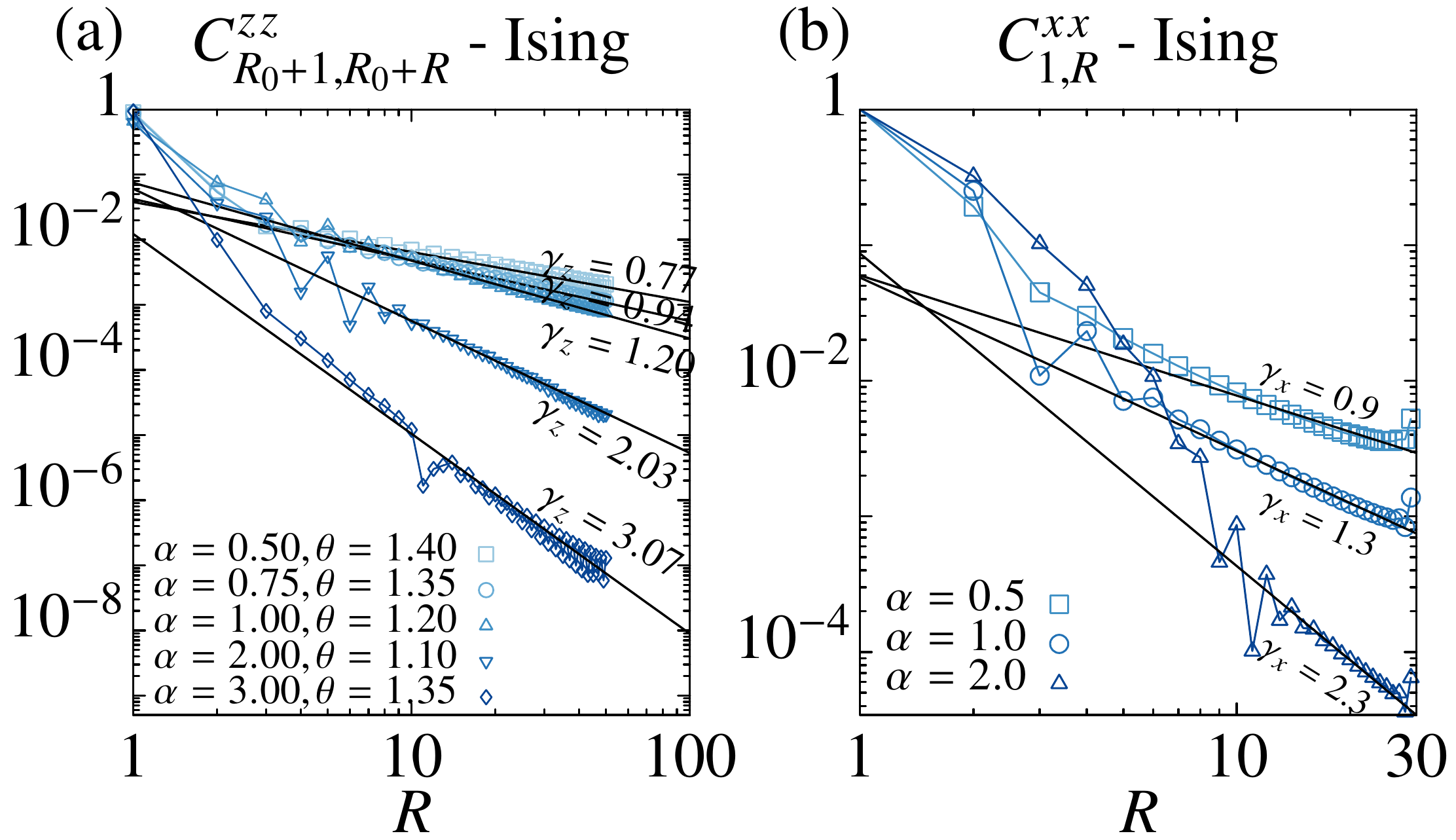}
\caption{(a): $C^{zz}_{R_0+1,R_0+R}$ correlation for the LRI for $\theta=0.35\pi$ (AM phase) and $L=200$, computed from $R_0=100$. The data are fitted by the solid lines $\sim 1/R^{\gamma_z}$, $\gamma_z$ reported in the plot. (b): $C^{xx}_{1,R}$ correlation at $\theta=0.207\pi$ (PM phases) for a system of $L=30$ sites. }
\label{fig:ZZIsing}
\end{figure}
In the AM phase, instead, we find that the decay of $C^{zz}_{i,j}$ [Fig.~\ref{fig:ZZIsing}(a)] for $\alpha \gtrsim 1$ displays an algebraic tail with an exponent compatible with $\gamma_z = \alpha$, mimicking $C^{xx}_{i,j}$ in the PM phase (a very precise estimate is forbidden by the decay oscillations between even and odd sites). These results are consistent with those of Ref.~\cite{Deng2005}, obtained for the case $\alpha=3$. Notably the decay exponent is here always different  from the value $2 \alpha$ analytically computed for $g_2(i,j) $ in Eq.~\eqref{eqn:DensDensKitaev}. This discrepancy is again probably due to the  role of the string operators, here quantitatively relevant even for $\alpha \gtrsim 1$. For $\alpha \lesssim 1$ again no universal behaviour for $\gamma_z$ is found.
\section{Edge modes properties}
\label{sec:Edges}
\subsection{Massless edge modes} \label{sec:EdgesA} \label{sec:MajoranaModes}

It is known that the SR Kitaev chain hosts modes localized at the edges~\cite{Kitaev2001} in the ordered phase for $\pi/4<\theta <3\pi/4$. These modes are fermionic and massless (in the limit $L \to \infty$), thus they have Majorana nature~\cite{Wilczek2009} and are a consequence of the topological non-triviality of the ordered phase (see e.g.~\cite{Fidkowski2011} and references therein).

\begin{figure}\centering
\includegraphics[width=0.5\textwidth-4pt]{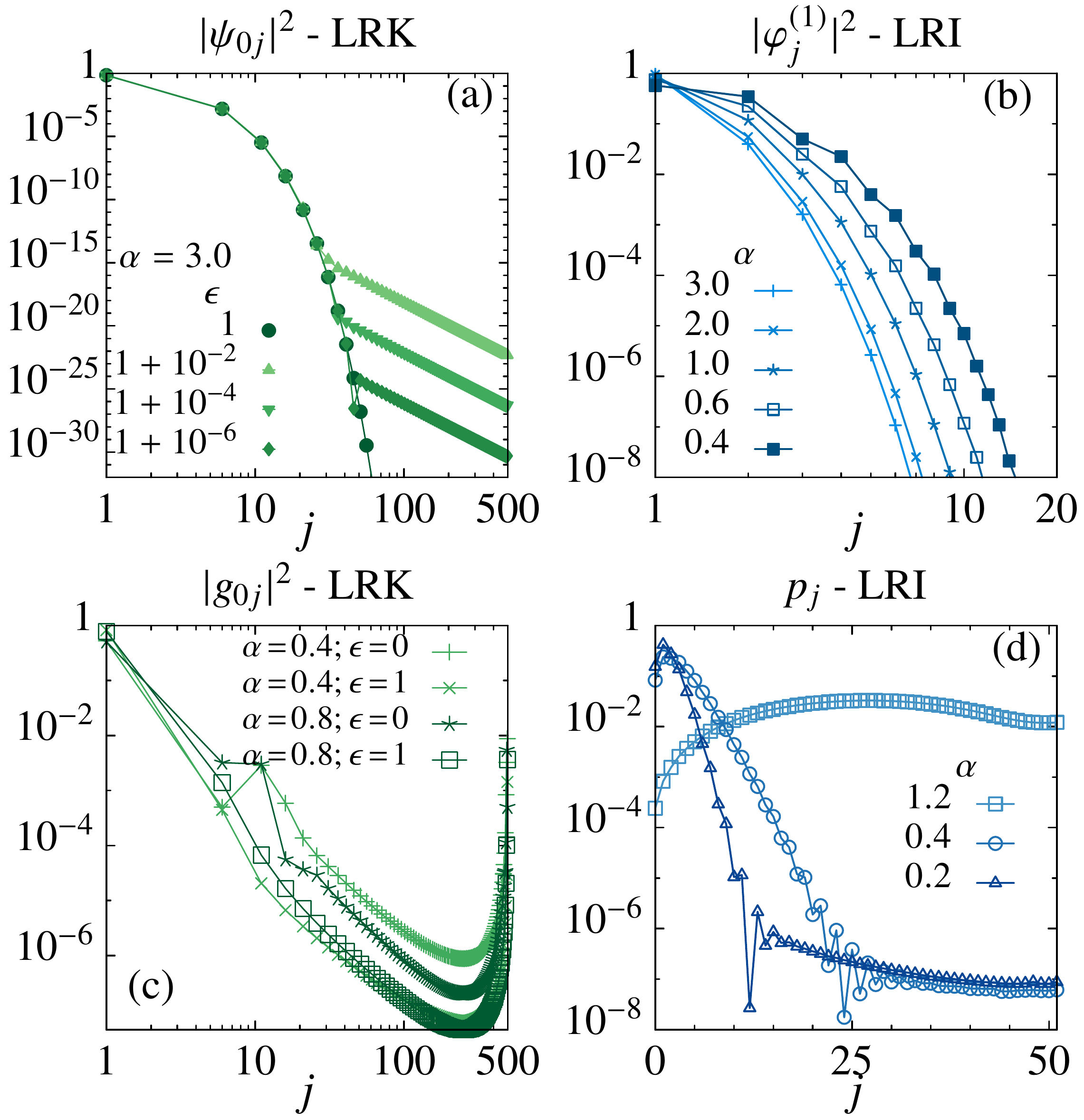}
\caption{ (a) LRK model: spatial probability distribution for the Majorana edge mode $\psi_{0j}$ in the AM1 phase for $\theta = 0.35\pi$, $\alpha=3$  and different $\epsilon$. When $\epsilon=1$, the mode is exponentially localized at one edge, while a power-law tail gradually appears when $\epsilon\neq 1$. (b) LRI model: spatial probability distribution for the mode $\varphi_j^{(1)}$ as function of the lattice site $j$ in the AM phase for different $\alpha$ and $\theta=0.45\pi$. The decay with $j$ is numerically found to be exponential. (c) LRK model: Spatial probability distribution for the massive edge mode in the AM2 phase for $\theta = 0.7\pi$ for different $\alpha$ and $\epsilon$. The decay with distance is purely power law and the probability distribution is symmetric with respect to half of the chain. (d) LRI model: Probability density $p_j=\abs{w^{(1)}_j}^2+\abs{w^{(2)}_j}^2$ of the two degenerate excited states $\ket{1^-}$ and $\ket{2^-}$ in the PM2 phase. When $\alpha\gtrsim 1$, $p_j$ spreads into the bulk of the chain. When $\alpha\lesssim 1$, $p_j$ is exponentially localized near a single edge. In panel (b) and (d), we plot only the points compatible with the DRMG errors.}
\label{edgeDLRK}
\end{figure}

\begin{figure*}\centering
\includegraphics[width=0.9\textwidth]{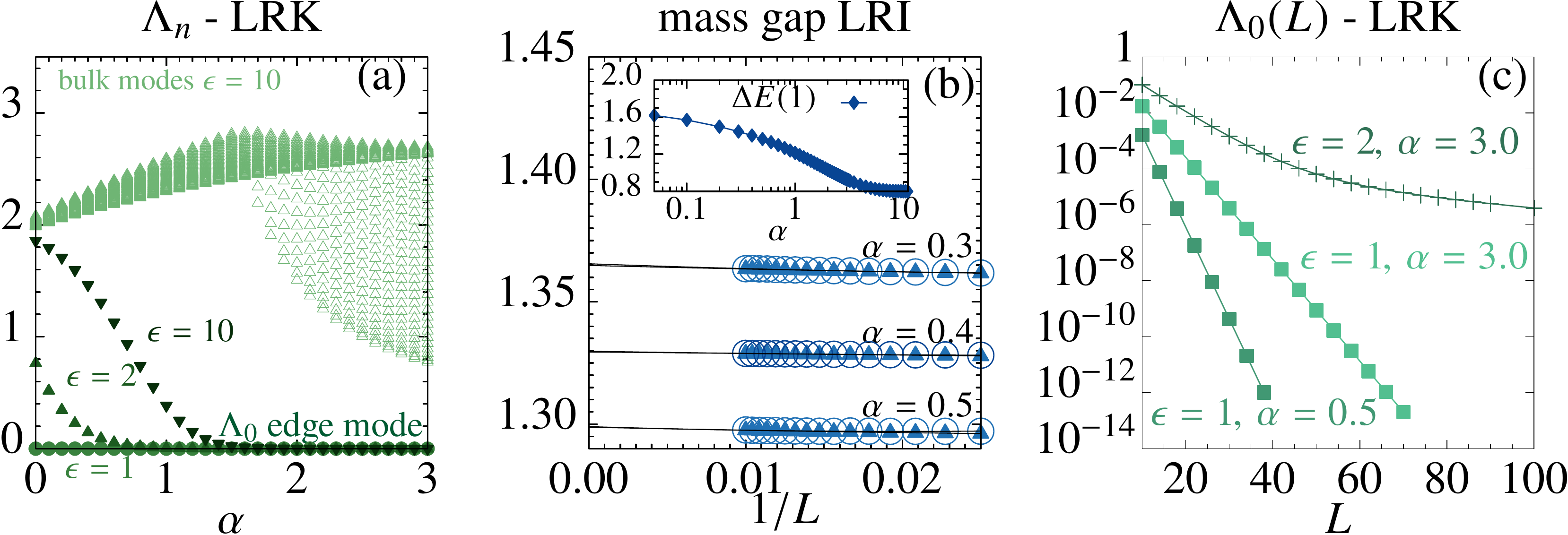}
\caption{(a) First 45 low energy states of the LRK models for $L=500$ and $\theta = 0.7\pi$ as function of $\alpha$. It is possible to distinguish the $\Lambda_0$ mode (solid points for three different $\epsilon$) separated from the $\Lambda_{n>0}$ bulk modes (light green triangles for $\epsilon=10$). When $\alpha\gtrsim 1$ the vanishing of $\Lambda_0$ for $\epsilon\neq 1$ signals the spontaneous breaking of the $\mathds{Z}_2$ symmetry of LRK and defines the AM1 phase [see also the phase diagram in Fig.~\ref{fig1}(b)]. When $\Lambda_0>0$ the $\mathds{Z}_2$ symmetry is restored and the phase AM2 appears. (b) Main panel:  $\Delta E (1)$ (empty circles) and $\Delta E (2)$ (solid triangles) for the LRI as functions of the inverse of the length of the chain $L$ for $\theta=0.159\pi$ and for three $\alpha$ reported in the plot. These two states are degenerate in the $L\to\infty$ limit, as finite-size scaling shows. Inset:   $\Delta E (1)$ as function of $\alpha$ for $\theta=0.159\pi$. When $\alpha$ decreases, $\Delta E (1)$ increases, showing a change for the energy spectrum in the region PM2 of the phase diagram of Fig.~\ref{fig1}(a). (c) Edge gap $\Lambda_0$ for (squares) $\epsilon=0.5$ and two different $\alpha$ as function of the system size $L$. When  $\epsilon=1$, the edge gap scales exponentially to zero for all $\alpha$. For comparison, the case (crosses) $\epsilon=2$ and $\alpha=3$ is also reported. There the edge mass displays an hybrid exponential and power-law decay.}
\label{DKLRmassive}
\end{figure*}

For the LRK models of Eq.~\eqref{double}, Majorana massless modes are found for $\alpha \gtrsim 1$ in the AM1 region of the phase diagram of Fig.~\ref{fig1}(b). Plotting  the square of the wavefunction $\psi_{0j}$ corresponding to the zero edge gap $\Lambda_0$, defined in Sec.~\ref{sec:edgeStates}, we find for $\epsilon \neq 1$ in Eq.~\eqref{double} a hybrid exponential and algebraic decay with the distance from one edge of the chain. 
The exponent of the algebraic decay of $\abs{\psi_{0j}}^2$ is found to be equal to $2\alpha$. This behaviour is similar to that observed in Ref.~\cite{Vodola2014} in the presence of LR pairing only. 

Interestingly, here the algebraic tail of the Majorana modes can be tuned to completely disappear by changing  the parameter $\epsilon$ that fixes the unbalance  between the hopping and the pairing terms in the Hamiltonian of Eq.~\eqref{double}. Figure~\ref{edgeDLRK}(a) shows an example for $\theta=0.4\pi$ and $\alpha=3$. When $\epsilon \neq 1$  the hybrid behaviour is fully visible and the state $\psi_{0j}$ decays in the bulk with an algebraic tail. However, by approaching the value $\epsilon=1$ the algebraic tail of $\psi_{0j}$ decreases and eventually disappears. As a result, for $\epsilon=1$ the wave function  $\psi_{0j}$ becomes exponentially localized at one edge. We find that this exponential localization for $\epsilon=1$ is present also in the parameter region $\alpha<1$. Moreover, the edge gap $\Lambda_0$ scales to zero exponentially with the system size $L$ for all $\alpha$ as Fig.~\ref{DKLRmassive}(c) shows.\\

For the LR Ising Hamiltonian of Eq.~\eqref{LRI}, edge modes appear in the AM phase for every $\alpha>0$. They have zero mass, since the edge gap $E_{\ket{0^{-}}} - E_{\ket{0^{+}}}$ [defined in Sec.~\ref{sec:edgeStates}] vanishes. Examples of these edge modes for different $\alpha$ are given in Fig.~\ref{edgeDLRK}(b), where we plot the square of  the wavefunction $\varphi^{(1)}_j$ defined in Eq.~\eqref{eqn:MajoranaLeft}.  We find numerically that $\varphi^{(1)}_j$ decays exponentially with the distance from the edge of the chain for all values of $\alpha$, as well as $\varphi^{(2)}_j$ (not shown) from the opposite edge.

\subsection{Massive edge modes}\label{sec:EdgesB} 
If we extend the analysis of the LR Kitaev Hamiltonian of Eq.~\eqref{double} to different $\epsilon$ and sufficiently small $\alpha$, a totally new situation arises for the edge gap $\Lambda_0$ and the edge modes:
In the region denoted as AM2 in the phase diagram of Fig.~\ref{fig1}(b), we find that  $\Lambda_0$ (which is zero for $\alpha\gtrsim 1$) becomes nonzero for $\alpha \lesssim 1$ and $\epsilon\neq 1$, also in the thermodynamical limit.

This  case is shown in Fig.\ref{fig1}(e), where we plot the edge gap $\Lambda_0$ as function of $\alpha$ and $\theta$ for $\epsilon=10$.
Between $\epsilon = 1$ and $\epsilon = 10$ we checked a continuous increase of the extension of the region AM2 and no transitions in between.
Similarly in Fig.~\ref{DKLRmassive}(a) we plot $\Lambda_0$ together with several other single-particle energies as a function of $\alpha$ and for a fixed $\theta$. These energies have been computed as described in Sec.~\ref{sec:edgeStates}. In Fig.~\ref{DKLRmassive}(a) the mass of the edge mode $\Lambda_0$ is easily recognizable for all $\epsilon$, since it is separated from all bulk modes by a finite gap. \\

Consistently with the discussion above, for $\alpha \gtrsim 1$, $\Lambda_0$ is zero as expected from the SR model, so that  two degenerate ground states exist as the $\mathds{Z}_2$ symmetry of the model is spontaneously broken. However, surprisingly, for $\alpha \lesssim 1$ and $\epsilon\neq 1$, we find that $\Lambda_0$ becomes finite and thus the ground state is unique. This indicates that the  $\mathds{Z}_2$ symmetry of the model (which is broken for $\alpha \gtrsim1$ in the AM1 phase) is restored for $\alpha \lesssim 1$. As a consequence, the region AM2, where the $\mathds{Z}_2$ symmetry is restored,  must be separated from AM1 by a quantum phase transition, even if no closure for the mass gap arises in the bulk. 

The wavefunction of the lowest massive $\Lambda_0$ state is  now given by the matrix element $g_{0i}$ defined in Sec.~\ref{sec:edgeStates}.
By plotting the probability density $\abs{g_{0i}}^2$, we now obtain a localization on the edges that is symmetric with respect to the middle of the chain.  This probability density decays algebraically when approaching half of the chain, as is clearly seen in Fig.~\ref{edgeDLRK}(d). \\

A similar wavefunction localization at the edges of the system is found also for the LRI model in the PM2 region in the phase diagram of Fig.~\ref{fig1}(a) for $\alpha \lesssim 1$, where the $\mathds{Z}_2$ symmetry is preserved.
However, while for the LRK massive edge modes originate from Majorana edge modes present at $\alpha \gtrsim 1$, for the LRI  the edge localization arises for excited states of the bulk spectrum in the region PM1.
 For $\alpha\lesssim 1 $, these states are degenerate, as shown in Fig.~\ref{DKLRmassive}(b), and separated from the third excited state by a gap that is finite in the thermodynamic limit. Because of this degeneracy, we consider the probability density $p_j=\abs{w^{(1)}_j}^2+\abs{w^{(2)}_j}^2$, with $w^{(1,2)}_j$ wavefunctions of $\ket{(1,2)}$ defined in Sec.~\ref{sec:edgeStates}. A typical situation is depicted in Fig. \ref{edgeDLRK}(d), where we plot $p_j$ as function of the lattice site $j$ for different values of $\alpha$. For $\alpha\gtrsim 1 $, $p_j$ is oscillating  and delocalized in the bulk, while it is localized exponentially at the edges for $\alpha\lesssim 1$ and is symmetric with respect to half of the chain.\\

We leave as an open question whether the edge localization here signals the appearance of a new phase  (and without mass-gap closure) with preserved $\mathds{Z}_2$ symmetry, similar to the LRK models above.

\section{Observability in current experiments} \label{sec:CorrD}
Recent experiments with cold ions have made possible the realization of  LR Ising-type Hamiltonians as Eq.~\eqref{LRI} with $0\lesssim \alpha \lesssim 3.5$~\cite{Islam2013, Britton2012, Richerme2014,Jurcevic2014}. In these experiments, both static and dynamical spin-spin correlations, as well as the spectrum of quasi-particle excitations~\cite{Jurcevic2015}, can be measured with extreme precision, which in principle  could allow for an analysis of some of the observables discussed above. For example, the observation of long-distance algebraic of correlations, as well as spectroscopic signatures of the formation of localized excited edge modes for $\alpha\lesssim 1$ could allow for the precise determination of the properties of these LR models. 

One key aspect of experiments, however, is that experimentally attainable lengths for ion chains are currently limited to at most few tens of ions.  It is thus natural to ask whether the characteristic long-distance decay of correlation functions described above can be observed in systems of such length. To explore this issue, Fig.~\ref{fig:ZZIsing} (right panel) shows the correlation $C^{xx}_{1,R}$ for Hamiltonian Eq.~\eqref{LRI} in the PM1 and PM2 phases for a chain of $L=30$ sites with open boundary conditions and for different $\alpha$. For $\alpha>2$, the initial exponential decay dominates the correlations for $R\lesssim 10$, while a comparatively small algebraic tail is found for $R\gtrsim 10$. For $\alpha \lesssim 1$, however, the exponential part has essentially disappeared and the decay is purely algebraic at all distances, as expected from the discussion of Secs.~\ref{sec:CorrC}. This fundamental change of behaviour  around $\alpha\simeq 1$ may be observable. We note, however, that the exponent $\gamma_x$ of the algebraic decay is here different from that presented in Fig.~\ref{fig:EEXX}(a), due to strong finite size effects in these systems. We find similar results for the correlation $C^{zz}_{1,R}$.

On the other hand, the emergence of massive edge modes in the LRI chain could be a convenient diagnostic of the change of nature of the paramagnetic phase for $\alpha\lesssim 1$.\\

\section{Summary and outlook}\label{Conclusions}

In this work we have analyzed the phase diagram of the long range anti-ferromagnetic Ising chain and of a class of fermionic Hamiltonian of the Kitaev type, with long-range pairing and hopping. We have clarified in what regions of the phase diagram violation of the area law occurs, and have provided numerical evidence and exact analytical results for the observed hybrid decay of correlation functions, which are found to decay exponentially at short range and algebraically at long range, for all $\alpha$. We have further demonstrated the breaking of conformal symmetry along the critical lines in both models  at low enough $\alpha$. Most interestingly, for the fermionic models we have demonstrated for the first time that the topological edge modes can become massive for sufficiently small values of $\alpha\lesssim 1$. This implies the existence of a transition to a novel phase without closure of the mass gap, to the known phase with massless Majorana modes for $\alpha\gtrsim 1$.
We conjecture that the possibility of a phase transition with nonzero mass gap is due to the peculiar behaviour of LR correlations, showing power-law tails also when the gap does not vanishes.
 Similarly, we have found that excited bulk states in the paramagnetic phase of the Ising model can become localized at the edges of the chain for $\alpha\lesssim 1$.\\

This works may open several exciting research directions. The first question concerns the nature and topological properties of the proposed new phase of the Kitaev model with $\alpha\lesssim 1$, and of its localized edge modes.  We conjecture that these massive edge modes are due to the hybridization of the Majorana modes at small $\alpha$, due to the bulk overlap between their wave functions, whose decay is slower and slower for decreasing $\alpha$. This aspect will be the subject of future studies. \\
Another important open question is  whether  the appearance of massive edge modes may be connected also to the violation of the area-law for the entanglement entropy in these models. 

In general, these results represent counter-examples for the topological properties of existing topological models with long-range interactions, as recently analyzed in~\cite{Gong2015}. The question of whether a possible universal behaviour exists for topological models with long-range interactions is thus still wide open.

\acknowledgments{
We are pleased to thank Alessio Celi, Domenico Giuliano, Alexey Gorshkov, Miguel Angel Martin-Delgado, Andrea Trombettoni and Oscar Viyuela Garcia for fruitful discussions and Fabio Ortolani for help with the DMRG code.
We acknowledge support by the ERC-St Grant ColdSIM (No. 307688), EOARD, UdS via Labex NIE and IdEX,  RYSQ, computing time at HPC-UdS.
}

\end{document}